\documentclass[final]{IEEEtran}
\usepackage{amsmath}
\usepackage{amsthm}
\usepackage{multirow}
\usepackage{color,cite}
\usepackage{subfigure}
\usepackage{mathrsfs}
\usepackage{amssymb}
\usepackage{bm}
\usepackage{graphicx}
\usepackage{amsfonts}
\usepackage{algorithm}
\usepackage{algorithmic}
\usepackage{makecell}
\usepackage{etoolbox}

\newtheorem{theorem}{Theorem}

\newtheorem{proposition}{Proposition}
\newtheorem{definition}{Definition}

\allowdisplaybreaks[4]

\input epsf

\begin{document}

\title{Near Interference-Free Space-Time User Scheduling for MmWave Cellular Network}

\author{Ziyuan Sha, Siyu Chen, and Zhaocheng Wang,~\emph{Fellow, IEEE}

\thanks{This work was supported by the National Key R\&D Program of China under Grant 2018YFB1801102. \emph{(Corresponding author: Zhaocheng Wang.)}}
\thanks{Z.~Sha, S.~Chen, and Z.~Wang are with Beijing National Research Center for Information Science and Technology, Department of Electronic Engineering, Tsinghua University, Beijing 100084, China, and Z. Wang is also with the Tsinghua Shenzhen International Graduate School, Shenzhen 518055, China. (E-mails: shazy17@mails.tsinghua.edu.cn, chensy18@mails.tsinghua.edu.cn, zcwang@tsinghua.edu.cn).}
}

\maketitle

\begin{abstract}
 The highly directional beams applied in millimeter wave (mmWave) cellular networks make it possible to achieve \emph{near interference-free} (NIF) transmission under judiciously designed space-time user scheduling, where the power of intra-/inter-cell interference between any two users is below a predefined threshold. In this paper, we investigate two aspects of the NIF space-time user scheduling in a multi-cell mmWave network with multi-RF-chain base stations. Firstly, given that each user has a requirement on the number of space-time resource elements, we study the NIF user scheduling problem to minimize the unfulfilled user requirements, so that the space-time resources can be utilized most efficiently and meanwhile all strong interferences are avoided. A near-optimal scheduling algorithm is proposed with performance close to the lower bound of unfulfilled requirements. Secondly, we study the joint NIF user scheduling and power allocation problem to minimize the total transmit power under the constraint of  rate requirements. Based on our proposed NIF scheduling, an energy-efficient joint scheduling and power allocation scheme is designed with limited channel state information, which outperforms the existing independent set based schemes, and has near-optimal performance as well.
\end{abstract}


\IEEEpeerreviewmaketitle

\section{Introduction}\label{sec_intro}
 The analog beamforming technique is widely adopted in millimeter wave (mmWave) communications \cite{background1, background2, background3, background4} to generate highly directional beam which mitigates the large path loss in mmWave band. By connecting one radio frequency (RF) chain to multiple antenna elements through a phase shifter array, the beam direction of that RF chain could be controlled via configuring the beamforming vector of the phase shifter array. To facilitate the practical implementation of analog beamforming, the beamforming vector for each user is usually selected from a codebook via a beam training procedure \cite{background4, background5, background6}, ensuring that each user is served by the beam with maximum directional power gain.

 Due to the application of highly directional beam, the interference characteristics in mmWave cellular network \cite{interference1, interference2, interference3} are different from the traditional RF cellular network \cite{low1, low2} with relatively wide radiation pattern. Specifically, in mmWave band, the downlink interference from a base station (BS) to a user could be strong only when the beam direction of the BS points towards the user, which implies that the strong interference between two users requires a special spatial relationship, e.g. co-located users. As a result, the overall interference level in mmWave network is significantly reduced in spatial domain \cite{interference1}, where only a small proportion of users are interference-limited if the BS deployment is not dense \cite{interference3}. Furthermore, for several users interfering with each other, we can schedule them in a time division multiple access (TDMA) manner \cite{mmWave1, mmWave2, mmWave3, mmWave4, mmWave5, mmWave6, mmWave0} so that any two users would not be served simultaneously and hence the interference is avoided.

 Therefore, different from the traditional RF band, it is possible to completely avoid the strong interferences in mmWave network via a judiciously designed space-time user scheduling strategy \cite{mmWave3, mmWave4, mmWave5}. In this paper, we refer to the space-time user scheduling where the powers of all interferences are below a threshold as \emph{near interference-free} (NIF) scheduling. A remarkable benefit of NIF scheduling lies in its high energy efficiency. Since the power of intra-/inter-cell interference is low, a relatively low Tx power of BS could achieve a considerable signal-to-interference-plus-noise ratio (SINR) to fulfill the rate requirements of users. Moreover, as demonstrated in \cite{mmWave1}, the transmit power consumption of serving one user can be further reduced if more time resources are allocated to the user.

 Against the above background, our research focuses on the following two problems. The first one is how to efficiently utilize the network space-time resources in NIF scheduling to fulfill the user requirements on the number of space-time resource elements. Furthermore, the second problem is how to minimize the total transmit power in NIF scheduling under the constraints of user data rate. By solving these two problems, an energy-efficient joint user scheduling and power allocation scheme could be given.

\subsection{Related Works}\label{sec_intro_related}
 As summarized in \cite{low1, low2}, the interference management in the traditional RF network has been studied by many researches from different perspectives, including the multi-antenna signal processing techniques in physical layer, e.g. \cite{low3, low4}, the coordinated user scheduling or power allocation in media access control (MAC) layer, e.g. \cite{low5, low6}, and the joint beamforming and user scheduling optimization, e.g. \cite{low7, low8}. In traditional RF band, the network performance is commonly optimized in consideration of the strong interference \cite{low5, low6, low7, low8}. However, the directional beam in mmWave band allows us to perform NIF user scheduling \cite{mmWave1, mmWave3, mmWave4, mmWave5, mmWave7} with very low interference level. As a result, the spirit of NIF user scheduling in mmWave network becomes different from the scheduling in traditional RF network.

 In recent years, a number of researches have been conducted on the user scheduling problem in multi-cell mmWave network, including \cite{mmWave1, mmWave2, mmWave3, mmWave4, mmWave5, mmWave7, mmWave8, mmWave9}. Under various scenarios and objective functions, time-domain user scheduling is studied in \cite{mmWave1, mmWave2, mmWave3, mmWave4, mmWave5, mmWave7}, and frequency-domain user scheduling is investigated in \cite{mmWave8, mmWave9}, where user association is considered together with user scheduling in \cite{mmWave2, mmWave3, mmWave7, mmWave9}. Specifically, \cite{mmWave1} minimizes the transmit power in heterogeneous network, and \cite{mmWave2} maximizes the sum rate in a multi-tier network with relaying access points. An interference-free scheduling scheme under max-min rate criterion is studied in \cite{mmWave3}. In our previous researches, we minimize the number of beam collisions caused by inter-cell interference in a two-cell multi-RF-chain scenario \cite{mmWave4} and a multi-cell single-RF-chain scenario \cite{mmWave5}, respectively. In \cite{mmWave7}, the proportional fair joint user association and time allocation problem is solved under interference-free assumption.

 Nevertheless, the NIF user scheduling in a general multi-cell mmWave network with multi-RF-chain BSs has not been studied in the literature. The existing studies \cite{mmWave1, mmWave3, mmWave5} related to NIF scheduling only consider single-RF-chain BS where the intra-cell interference does not exist, while \cite{mmWave4} investigates a relatively simple two-cell scenario with multi-RF-chain BSs, but the intra-cell interference remains untreated. However, it is common for mmWave BS to equip with multiple RF chains enabling spatial division multiple access (SDMA) of users \cite{mmWave01, mmWave02}, leading to more flexible user scheduling and non-negligible intra-cell interference as well. As a result, the NIF user scheduling in multi-cell network with multi-RF-chain BSs is a more challenging problem compared with the aforementioned existing studies.

 On the other hand, to facilitate NIF scheduling, interference graph is a widely adopted tool to characterize the strong interferences between users (or links) in the network \cite{mmWave1, mmWave5, IS1, IS2, IS3, IS4, IS5}. The independent set (IS) of interference graph is usually adopted to give the NIF user scheduling in a time slot, and the entire time-domain user scheduling can be derived by the combination of ISs. The schedule schemes in \cite{mmWave1, IS2, IS3} are based on maximum IS (MIS), and \cite{IS1, IS4, IS5} are based on maximum-weight IS (MWIS). However, in multi-cell network with multi-RF-chain BSs, each BS needs to select the users to be served at each time slot according to the number of RF chains. Consequently, the feasible ISs for each time slot would be too many to enumerate if the number of users becomes large. Meanwhile, the scheduling based on limited number of ISs \cite{mmWave1} is difficult to approach the optimal performance. Therefore, the existing IS based methods are insufficient to solve the NIF scheduling problem satisfactorily.

\subsection{Our Contributions}\label{sec_intro_contribution}
 The main contributions of our research are summarized as below.

\begin{itemize}
 \item We study the NIF user scheduling and power allocation in multi-cell mmWave network with multi-RF-chain BSs, where both intra-cell and inter-cell interferences are taken into account. The system model applied in our research is more general and challenging in comparison with the existing works.

 \item To fully utilize the space-time resources under NIF condition, we formulate the \textbf{NIF space-time scheduling problem} minimizing the unfulfilled user requirements on space-time resources, which is an unsolved problem in the literature. A near-optimal \textbf{NIF scheduling algorithm} is proposed to solve this problem. For the first time, we provide the insight that it is possible to design a space-time user scheduling which avoids all strong interferences and fulfills all user requirements at the same time.

 \item Based on the NIF space-time scheduling problem, we further study the \textbf{joint user scheduling and power allocation problem} minimizing the total transmit power, which has not been solved optimally or near-optimally in multi-cell network yet. We convert this problem into an integer convex optimization under NIF condition, and propose a near-optimal \textbf{joint user scheduling and power allocation scheme} accordingly, which outperforms its traditional counterparts.

 \item The analytical lower bounds of the NIF space-time scheduling problem and the joint user scheduling and power allocation problem are derived respectively. Our simulation results certify that the performance of our proposed NIF scheduling algorithm and joint user scheduling and power allocation scheme could both approach the lower bounds.
\end{itemize}

 The rest of this paper is organized as follow. Section \ref{sec_system} introduces our system model. Section \ref{sec_NIFSche} studies the space-time NIF scheduling problem.
 Based on our proposed NIF scheduling in Section \ref{sec_NIFSche}, we manage to solve the joint space-time user scheduling and power allocation problem in Section \ref{sec_SchePower}. Simulation results are provided in Section \ref{sec_sim} to evaluate the performance of our proposal. Section \ref{sec_conclusion} draws the conclusions.

\section{System Model}\label{sec_system}
 We study the downlink transmission of a $K$-cell mmWave cellular network where each cell has one serving BS. All users are considered to have accessed to its local BS already, so the association between users and BSs is fixed. Each BS is equipped with $N_{RF}$ RF chains and has $U$ single-RF-chain users to be served, where we usually have $N_{RF}<U$ because the number of RF chains in mmWave is limited due to hardware constraints \cite{background1}. In our system model, the numbers of users and RF chains, i.e. $U,N_{RF}$, are assumed to be the same for $K$ cells to simplify mathematical expressions, while the extension of our research to unequal numbers of users or RF chains over $K$ cells is straightforward.

 The time scale for one space-time user scheduling period consists of $N$ time slots, where a time slot is the minimum granularity for time-domain scheduling. We define one time slot used by a single RF chain as a space-time \emph{resource element} which can be assigned to serve one specific user. Therefore, the total number of the available resource elements in the network is $KN_{RF}N$ in each scheduling period. We consider that each user has a requirement on data rate to be satisfied, where the rate requirement of the $u$th user in cell $k$ is denoted as $\gamma_{k,u}$.

 We denote the space-time user scheduling of the $k$th cell as matrix $\bm{S}^k\in \mathbb{N}^{N_{RF}\times N}$, where $\mathbb{N}$ represents the set of non-negative integer. The element $s^k_{r,n}$ in the $r$th row and the $n$th column of $\bm{S}^k$ indicates that the $s^k_{r,n}$th user in cell $k$ is scheduled on the resource element of the $r$th RF chain at time slot $n$. If $s^k_{r,n}=0$, no user is scheduled on the corresponding resource element, leading to an unassigned resource element. Since all users are assumed to have single RF chain, the multi-stream space division multiplexing (SDM) transmission to one user is not considered, i.e. for any two non-zero $s^k_{r_1,n},s^k_{r_2,n}$, we have $s^k_{r_1,n}\ne s^k_{r_2,n}$ if $r_1\ne r_2$.

 Similarly, the power allocation of the $k$th cell is denoted by matrix $\bm{P}^k\in \mathbb{R}^{N_{RF}\times N}_+$, where $\mathbb{R}_+$ represents the set of non-negative real number, and the element $p^k_{r,n}$ in the $r$th row and the $n$th column of $\bm{P}^k$ represents the power allocated to the $r$th RF chain in time slot $n$. Obviously, we have $p^k_{r,n}=0$ if $s^k_{r,n}=0$.

\subsection{Beam Pattern}\label{sec_system_beam}

\begin{figure}[tp]
 \begin{center}
 \includegraphics[width=0.95\columnwidth]{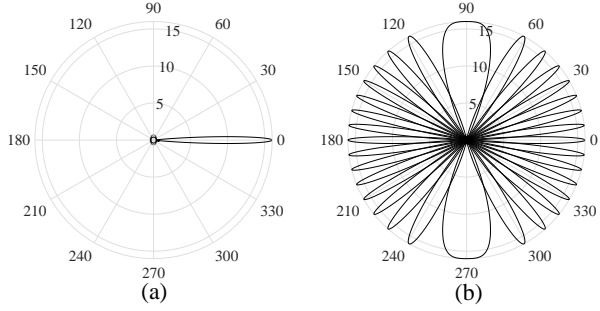}
 \end{center}
 \vspace{-5mm}
 \caption{(a) The beam pattern of a single beam with $N_t=16$, $b=0$ (the first beam), $g_{min}=0.4$; (b) The main lobes of $2N_t=32$ beams indexed in counter-clockwise order, covering all directions with unequal power gain from $\frac{4}{\pi^2}N_t$ to $N_t$.}
 \vspace{-5mm}
 \label{fig_beam}
\end{figure}

 We denote the number of antennas at the BS by $N_t$. As shown in Fig.~\ref{fig_beam}(a), the power gain of the main lobe of Tx beam is modeled based on the discrete fourier transform (DFT) codebook \cite{background4, interference1}, while the side-lobe power gain is approximated by a small constant $g_{min}$. The corresponding directional beamforming gain $G^T_b(\theta)$ with respect to azimuth angle $\theta$ of the $b$th Tx beam is expressed as
\begin{align*}
 G^T_b(\theta)=\left\{\begin{array}{cc}
 \frac{1}{N_t}\frac{|sin(\frac{N_t\pi}{2}(sin\theta-\frac{2f(b)}{N_t}))|^2}{|sin(\frac{\pi}{2}(sin\theta-\frac{2f(b)}{N_t}))|^2},\!\!\!&\theta\in[\theta_{b,1},\theta_{b,2}],\\
 g_{min}, &\text{otherwise},
 \end{array} \right.
\end{align*}
 where
\begin{align*}
 f(b)=\left\{\begin{array}{cc}
	N_t-b,\!\!\!&\frac{N_t}{2}\leq b\leq (\frac{3N_t}{2}-1),\\
	b, &\text{otherwise},
\end{array} \right.
\end{align*}
 and $[\theta_{b,1},\theta_{b,2}]$ represents the angular range of the main lobe of the $b$th beam, bounded by the two zero-gain angles of the main lobe. The beam index $b=0,1,\cdots,(2N_t-1)$ including $2N_t$ beams. Each user would be served by the Tx beam with maximum directional power gain. As illustrated in Fig.~\ref{fig_beam}(b), the $2N_t$ Tx beams cover all directions in $[0,2\pi]$ with unequal power gain from $\frac{4}{\pi^2}N_t$ to $N_t$. Compared with the relatively simple flat-top beam pattern \cite{mmWave2, mmWave3, mmWave4, mmWave5}, the main lobes of two neighboring Tx beams in our model partially overlap with varying gain, which could better reflect the possible strong intra-cell interference between nearby users.

 On the other hand, since the beam pattern of users is generally wide and different user equipments may have different Rx beam patterns, we consider the worst case for the user side that omni-directional Rx beam pattern is adopted \cite{mmWave5, mmWave8}, where the directional beamforming gain is $G^R(\theta)=1$. Nevertheless, we point out that our proposed scheduling scheme does not rely on the specific model of the beam pattern at the user or BS side.

\subsection{SINR and Rate Definitions}\label{sec_system_signal}
 We denote the path loss from the $k$th BS to the $u$th user served by the $k'$th BS as $L^k_{k',u}$. The power of the received signal from the $r$th RF chain of the $k$th BS to the $u$th user served by the $k'$th BS at time slot $n$ can be expressed as $p^k_{r,n}L^k_{k',u}G^{k,r,n,T}_{k',u}G^{k,R}_{k',u}$, where $G^{k,r,n,T}_{k',u}$ denotes the Tx beamforming gain of the beam of the $k$th BS serving the $s^k_{r,n}$th user to the $u$th user served by the $k'$th BS, and $G^{k,R}_{k',u}$ represents the Rx beamforming gain from the $k$th BS to the $u$th user served by the $k'$th BS. Since we consider $G^{k,R}_{k',u}=1$ as discussed in Section \ref{sec_system_beam}, $G^{k,R}_{k',u}$ will be omitted in the following discussions.

\begin{figure*}[tp]
 \begin{align}\label{equ_SINR}
  \rho^k_{r,n}=\frac{p^k_{r,n}L^k_{k,s^k_{r,n}}G^{k,r,n,T}_{k,s^k_{r,n}}}{p_n+\underbrace{\sum_{r',r'\ne r} p^k_{r',n}L^k_{k,s^k_{r,n}}G^{k,r',n,T}_{k,s^k_{r,n}}}_{\text{intra-cell interference}}
  +\underbrace{\sum_{k',k'\ne k}\sum_{r'} p^{k'}_{r',n}L^{k'}_{k,s^k_{r,n}}G^{k',r',n,T}_{k,s^k_{r,n}}}_{\text{inter-cell interference}}},
 \end{align}
 \hrulefill
 \vspace*{-3mm}
\end{figure*}

 Then, the SINR $\rho^k_{r,n}$ of the user scheduled on the resource element of the $r$th RF chain in the $k$th BS at time slot $n$ is given by (\ref{equ_SINR}) at the top of next page, where $p_n$ denotes the noise power. Note that the SINR in (\ref{equ_SINR}) is without Tx digital precoding, while the scenario with Tx precoding will be further discussed in Section \ref{sec_SchePower_CSI}. The average rate $r_{k,u}$ of the $u$th user in the $k$th cell over the $N$ time slots in one scheduling period is
\begin{align}\label{equ_averRate}
 r_{k,u}=\frac{1}{N}\sum_{(r,n),s^k_{r,n}=u}W\text{log}_2(1+\rho^k_{r,n}),
\end{align}
 where $W$ represents the system bandwidth, and all BSs use full $W$ bandwidth.

\subsection{Interference Graph}\label{sec_system_graph}
\begin{figure}[tp]
 \begin{center}
 \includegraphics[width=\columnwidth]{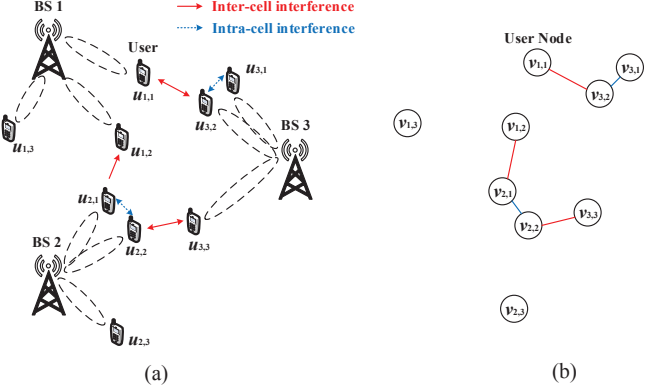}
 \end{center}
 \vspace{-5mm}
 \caption{(a) An example of a $3$-cell network serving $9$ users; (b) The corresponding interference graph.}
 \vspace{-5mm}
 \label{fig_network}
\end{figure}

 Since the highly directional beam pattern makes the interference between two different users be either strong or very weak, interference graph is a natural tool to represent the strong interference between any two users. As illustrated in Fig.~\ref{fig_network}(a), there might be strong intra- or inter-cell interference between two users in the network. Therefore, we model the network as an undirected graph $G=\{V,E\}$, where users are represented by the node set $V$, and the edge set $E$ represents the interferences between users, as shown in Fig.~\ref{fig_network}(b). Specifically, the $u$th user served by the $k$th BS is denoted by node $v_{k,u}\in V$. If there exists interference between the two users $v_{k,u}$ and $v_{k',u'}$, we have an edge $e_{(k,u),(k',u')}\in E$ connecting $v_{k,u}$ and $v_{k',u'}$, indicating that users $v_{k,u}$ and $v_{k',u'}$ should not be served simultaneously. Clearly, the interference graph in mmWave network is sparse due to the narrow beam, which facilitates the NIF space-time user scheduling.

 To decide whether the interference between two users is strong enough to be included in $G$, we set a predefined threshold $\epsilon$ for the ratio of the interference power to the desired signal power.
 Since frequent update of $G$ could induce high measurement overhead, all BSs are assumed to transmit reference signal with fixed power $P_0$ for interference measurement in the establishment of $G$, so that $G$ is independent of power allocation and does not need to update frequently.
 For any two users $v_{k,u}$ and $v_{k',u'}$, we define $G^{k,u}_{k,u}$ as the local beamforming gain from BS $k$ to user $v_{k,u}$, and $G^{k,u}_{k',u'}$ as the beamforming gain from the beam used to serve user $u$ by BS $k$ to the user $u'$ served by BS $k'$. In accord with the beamforming gain defined in Section \ref{sec_system_signal}, we have $G^{k,u}_{k,u}=G^{k,r,n,T}_{k,u}$ and $G^{k,u}_{k',u'}=G^{k,r,n,T}_{k',u'}$ if $s^k_{r,n}=u$. The desired power of users $v_{k,u}$ and $v_{k',u'}$ from their local BSs are $P_0L^k_{k,u}G^{k,u}_{k,u}$, $P_0L^{k'}_{k',u'}G^{k',u'}_{k',u'}$ respectively, and the interference power from the desired signal of $v_{k,u}$ to $v_{k',u'}$ and from the desired signal of $v_{k',u'}$ to $v_{k,u}$ are $P_0L^k_{k',u'}G^{k,u}_{k',u'}$, $P_0L^{k'}_{k,u}G^{k',u'}_{k,u}$ respectively. Then, we have an edge $e_{(k,u),(k',u')}\in E$ between $v_{k,u}$ and $v_{k',u'}$, if
\begin{align}\label{equ_threshold}
 &(L^{k'}_{k,u}G^{k',u'}_{k,u})/(L^k_{k,u}G^{k,u}_{k,u})>\epsilon,\\ \nonumber
 \text{or}~~&(L^k_{k',u'}G^{k,u}_{k',u'})/(L^{k'}_{k',u'}G^{k',u'}_{k',u'})>\epsilon.
\end{align}
 According to (\ref{equ_threshold}), the edge between two user nodes is determined by whether the desired signal of one user would cause a non-negligible interference on the other, where the path loss terms may result from either line-of-sight (LoS) channel or non-LoS (NLoS) channel. Note that a small $\epsilon$ could take more interferences into account, but makes the interference graph less sparse, so it is important to choose a proper $\epsilon$ (see Section \ref{sec_NIFSche_illu}).

 Based on interference graph, a definition of NIF scheduling is given as below, implying that there is no interference exceeding the threshold in (\ref{equ_threshold}). Although the weak interferences below the threshold are not considered in NIF scheduling, it is shown in \cite{Proofs} that the sum power of weak interferences on each user is still weak due to the highly directional beam.
\begin{definition}\label{def_NIF}
 A scheduling $\{\bm{S}^k\}_{k=1}^{K}$ is NIF, if at any time slot $n$, for any two pairs of $(k,r)$, $(k',r')$ with $s^k_{r,n}s^{k'}_{r',n}\ne0$, we have $e_{(k,s^k_{r,n}),(k',s^{k'}_{r',n})}\notin E$.
\end{definition}

 Additionally, in practical systems, the interference graph could be established by measuring the reference signal (RS) of neighboring BSs. For example, in Fig.~\ref{fig_network}(a), BS $2$ assigns zero-power RS to user $u_{2,2}$ on which the RSs from BS $3$ or other RF chains of BS $2$ can be measured. According to the measured interference power feedback, the network can confirm the inter-cell interference between $v_{2,2}$ and $v_{3,3}$ and the intra-cell interference between $v_{2,2}$ and $v_{2,1}$ after several times of measurements. Since the interference graph is sparse and commonly slow time-varying, the overhead induced by the establishment of $G$ would be acceptable.

\section{Near Interference-Free Space-Time User Scheduling}\label{sec_NIFSche}
 Obviously, by letting some resource elements be unassigned, i.e. muted, we could always achieve NIF scheduling. However, the unassigned resource elements are wasted and lead to the sacrifice of data throughput to users. Therefore, a well-designed NIF scheduling is preferred to have no unassigned resource element, ensuring that the resource elements are fully utilized and meanwhile all strong interferences are avoided.

 In this section, we firstly consider the user requirement on the number of resource elements. We denote the requirement of user node $v_{k,u}$ on the number of resource elements as $d_{k,u}$, where $d_{k,u}$ could be derived from the user rate requirement $\gamma_{k,u}$ in consideration of channel quality and interference power. In Section \ref{sec_SchePower}, we will elaborate on the method to design $d_{k,u}$ which fulfills the user rate requirements $\gamma_{k,u}$ and minimizes the total transmit power.
 The sum requirements on resource elements of all users in one cell is $N_{RF}N$, which can be expressed as
\begin{align}\label{equ_dPerCell}
 \sum_{u} d_{k,u}=N_{RF}N,~~\forall k,
\end{align}
 indicating that all resource elements are allocated to the users in the cell. Therefore, our objective is to derive the NIF scheduling which minimizes the number of unfulfilled requirements on resource elements so that the space-time resources are utilized most efficiently.

 Hereby, we allow $d_{k,u}$ to be arbitrarily given but satisfy (\ref{equ_dPerCell}), so that our proposed NIF user scheduling algorithm in this section is applicable to the $d_{k,u}$ given for various purposes.

\subsection{Problem Formulation}\label{sec_NIFSche_problem}
 The total number $n_0$ of the unfulfilled requirements on resource elements of all users is
\begin{align}\label{equ_n0}
 n_0=\sum_{(k,u)}(d_{k,u}-\hat{d}_{k,u}),
\end{align}
 where $\hat{d}_{k,u}$ represents the actual number of the resource elements assigned to the user node $v_{k,u}$, which is expressed as
\begin{align}\label{equ_dConstraint}
 \hat{d}_{k,u}=\sum_{(r,n)}\delta(s^{k}_{r,n},u)\leq d_{k,u},
\end{align}
 where $\delta(x,y)$ is an indicator function with $\delta(x,y)=1$ if $x=y$, and $\delta(x,y)=0$ if $x\ne y$. Note that the constraint $\hat{d}_{k,u}\leq d_{k,u}$ is introduced in (\ref{equ_dConstraint}), which guarantees that the number of unfulfilled requirements $n_0$ is exactly the number of unassigned resource elements given by
\begin{align}\label{equ_n0Alt}
 n_0=\sum_{(k,r,n)}\delta(s^{k}_{r,n},0).
\end{align}

 As mentioned previously, the SDM transmission is not allowed since the users have single RF chain, which could be written as
\begin{align}\label{equ_sdmConstraint}
 \delta(s^k_{r,n},s^{k}_{r',n})=\delta(r,r'),~~\text{if}~s^k_{r,n}s^{k}_{r',n}\ne0,
\end{align}
 indicating that one user cannot be served by two RF chains simultaneously. Meanwhile, it also implies that each user can be scheduled on at most $N$ resource elements over $N$ time slots, so the requirement $d_{k,u}$ should satisfy $0\leq d_{k,u}\leq N$.

 Combining (\ref{equ_n0}), (\ref{equ_dConstraint}), (\ref{equ_sdmConstraint}) with the NIF condition given by Definition \ref{def_NIF}, the NIF space-time user scheduling problem \textbf{P1} is formulated as
\begin{align}\label{equ_P1}
 \textbf{P1:}~~&\min\limits_{\{\bm{S}^k\}_{k=1}^{K}}~n_0=\sum_{(k,u)}(d_{k,u}-\hat{d}_{k,u})\\ \nonumber \tag{\ref{equ_P1}a}
 \text{s.t.}~~&~s^{k}_{r,n}\in\{0,1,\cdots,U\},~\forall k,r,n\\ \nonumber \tag{\ref{equ_P1}b}
 &\hat{d}_{k,u}=\sum_{(r,n)}\delta(s^{k}_{r,n},u)\leq d_{k,u},~\forall k,u\\ \nonumber \tag{\ref{equ_P1}c}
 &\delta(s^k_{r,n},s^{k}_{r',n})=\delta(r,r'),~\text{if}~s^k_{r,n}s^{k}_{r',n}\ne0,~\forall k,r,r',n\\ \nonumber \tag{\ref{equ_P1}d}
 &e_{(k,s^k_{r,n}),(k',s^{k'}_{r',n})}\notin E,~\text{if}~s^k_{r,n}s^{k'}_{r',n}\ne0,~\forall k,r,k',r',n.
\end{align}
 It can be seen that \textbf{P1} is an integer programming problem on a graph. We can derive a lower bound of $n_0$ and propose a scheduling algorithm for \textbf{P1} with near-optimal performance.

\subsection{Lower Bound of $n_0$}\label{sec_NIFSche_bound}
 We denote the set of time slots during which $v_{k,u}$ is served as $T_{k,u}=\{n|\exists r,~\text{s.t.}~s^{k}_{r,n}=u\}$. Obviously, we have $|T_{k,u}|=\hat{d}_{k,u}$ according to (\ref{equ_dConstraint}) and (\ref{equ_sdmConstraint}).

\begin{figure}[tp]
 \begin{center}
 \includegraphics[width=0.9\columnwidth]{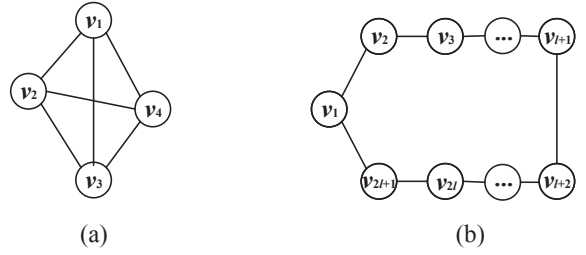}
 \end{center}
 \vspace{-5mm}
 \caption{(a) A clique with $4$ nodes; (b) An empty cycle with $2l+1$ nodes.}
  \vspace{-5mm}
 \label{fig_cliquecycle}
\end{figure}

 We consider two special structures in $G$ which may lead to unfulfilled requirements inevitably. The first structure is clique, i.e. a set of nodes where any two nodes are adjacent, as shown in Fig.~\ref{fig_cliquecycle}(a), and the second one is odd-length \emph{empty cycle}, i.e. an odd-length cycle where each node does not connect with other nodes in the cycle except its two neighbors, as shown in Fig.~\ref{fig_cliquecycle}(b). Especially, a length-$3$ empty cycle is also a $3$-node clique.

 The following two propositions can be derived for these two structures, respectively.
\begin{proposition}\label{prop_clique}
 For a $Q$-node clique with nodes $\{v_{k_q,u_q}\}_{q=1}^{Q}$, we have $\sum_{q}\hat{d}_{k_q,u_q}\leq N$.
\end{proposition}
\begin{proof}
 See \cite{Proofs}.
\end{proof}

 Consequently, we have $\sum_{q}(d_{k_q,u_q}-\hat{d}_{k_q,u_q})\geq \left(\sum_{q}d_{k_q,u_q}\right)-N$, which indicates that there would be at least $\left(\sum_{q}d_{k_q,u_q}\right)-N$ unfulfilled requirements in $n_0$ if $\sum_{q}d_{k_q,u_q}>N$.

\begin{proposition}\label{prop_cycle}
 For a length-$(2l+1)$ empty cycle with nodes $\{v_{k_i,u_i}\}_{i=1}^{2l+1}$, we have $\sum_{i}\hat{d}_{k_i,u_i}\leq lN$.
\end{proposition}
\begin{proof}
 See \cite{Proofs}.
\end{proof}

 Similarly, there would be at least $\left(\sum_{i}\hat{d}_{i}\right)-lN$ unfulfilled requirements in $n_0$ if $\sum_{i}\hat{d}_{i}>lN$.

 Therefore, by counting all the inevitable unfulfilled requirements caused by cliques and empty odd-length cycles, a lower bound of $n_0$ can be derived. However, as illustrated in Section \ref{sec_NIFSche_illu}, the empty cycles with $4$ or more nodes are very rare because there are usually inner edges in the cycle, which makes a cycle hard to be ``empty''. Meanwhile, the length-$3$ empty cycle is equivalent to a $3$-node clique. Therefore, we only consider the cliques in the derivation of the lower bound of $n_0$.

 Algorithm \ref{alg_bound} provides a method to compute the lower bound $\tilde{n}_0$ of $n_0$. In the algorithm, we repeat searching the local-maximal clique $G_c=\{V_c,E_c\}$ in $G$, i.e. the clique not contained in any other cliques, with $\sum_{v_{k_q,u_q}\in V_c} d_{k_q,u_q}>N$. If such a local-maximal clique is found, there would be $\sum_{v_{k_q,u_q}\in V_c} d_{k_q,u_q}-N$ inevitable unfulfilled requirements in $n_0$. Then, we remove $G_c$ and its connected edges from $G$ and try to find another target local-maximal clique, until no such clique can be found in $G$. We prove the validity of Algorithm \ref{alg_bound} in Proposition \ref{prop_bound}.

\begin{algorithm}
\caption{Derivation of the lower bound of $n_0$}
\label{alg_bound}
\begin{algorithmic}[1]
 \STATE Set $\tilde{n}_0=0$;
 \REPEAT
  \STATE Search for a local-maximal clique $G_c=\{V_c,E_c\}$ with $\sum_{v_{k_q,u_q}\in V_c} d_{k_q,u_q}>N$;
  \STATE Set $\tilde{n}_0=\tilde{n}_0+\left(\sum_{v_{k_q,u_q}\in V_c} d_{k_q,u_q}-N\right)$, and remove $G_c$ from $G$;
 \UNTIL no such clique can be found in $G$.
\end{algorithmic}
\end{algorithm}

\begin{proposition}\label{prop_bound}
 The optimal $n^*_0$ of \textbf{P1} with interference graph $G$ is not less than the $\tilde{n}_0$ given by Algorithm \ref{alg_bound}, i.e. $n^*_0\geq \tilde{n}_0$.
\end{proposition}
\begin{proof}
 See \cite{Proofs}.
\end{proof}

 It can be seen that $\tilde{n}_0$ is influenced by the user requirements $d_{k,u}$ significantly. Therefore, a judicious resource allocation $d_{k,u}$ should at least ensure that $\tilde{n}_0=0$, i.e. there is no clique such that the sum requirements of the user nodes in the clique exceeds $N$, which is a necessary condition to achieve NIF scheduling with zero unfulfilled requirement.

\subsection{RF-chain-sufficient Scheduling}\label{sec_NIFSche_sche1}
 To simplify the scheduling problem \textbf{P1}, we firstly consider a \emph{RF-chain-sufficient} scenario, i.e. each user is served by an exclusive RF chain. In this scenario, we only need to determine the time slots assigned to each user, i.e. $T_{k,u}$. In addition, the sum requirements of the users in each cell is still $N_{RF}N$ as shown in (\ref{equ_dPerCell}). The simplified problem \textbf{P2} can be expressed as
\begin{align}\label{equ_P2}
 \textbf{P2:}~~&\min\limits_{\{T_{k,u}\}_{k,u}}~n_0=\sum_{(k,u)}(d_{k,u}-|T_{k,u}|)\\ \nonumber \tag{\ref{equ_P2}a}
 \text{s.t.}~~&T_{k,u}\subset\{1,\cdots,N\},~\forall k,u\\ \nonumber \tag{\ref{equ_P2}b}
 &|T_{k,u}|\leq d_{k,u},~\forall k,u\\ \nonumber \tag{\ref{equ_P2}c}
 &T_{k,u}\cap T_{k',u'}=\emptyset,~\text{if}~e_{(k,u),(k',u')}\in E,~\forall k,u,k',u',
\end{align}
 where the constraints (\ref{equ_P2}a), (\ref{equ_P2}b), (\ref{equ_P2}c) correspond to the constraints (\ref{equ_P1}a), (\ref{equ_P1}b), (\ref{equ_P1}d) in \textbf{P1}, respectively. Then, we study \textbf{P2}
 before solving \textbf{P1}.

 Note that if $v_{k,u}$ is not connected with any other node in $G$, then $T_{k,u}$ can be derived by selecting any $d_{k,u}$ elements from $\{1,\cdots,N\}$, which does not violates the $3$ constraints of \textbf{P2} and induces zero unfulfilled requirement. Furthermore, since different connected components in $G$ are independent of each other, we can schedule the users in each connected component independently. Especially, an unconnected user node can be treated as a connected component with only one node.

\begin{algorithm}
\caption{RF-chain-sufficient Scheduling Algorithm}
\label{alg_sche1}
\begin{algorithmic}[1]
 \FOR{each connected component $G_i=\{V_i,E_i\}$ in $G$}
  \STATE Select one $v_{k_0,u_0}\in V_i$, and derive $T_{k_0,u_0}$ by selecting $d_{k_0,u_0}$ elements from $\{1,\cdots,N\}$;
  \STATE Set $V^t_i=\{v_{k_0,u_0}\}$;
  \WHILE{$V^t_i\ne V_i$}
   \STATE Search for the user nodes in $V_i\setminus V^t_i$ connected with $V^t_i$, from which select the user node $v_{k,u}$ with maximum number of connected nodes in $V^t_i$;
   \STATE Derive set $T'_{k,u}$ of the available time slot to schedule $v_{k,u}$, and derive $T_{k,u}$ by selecting $\text{min}\{|T'_{k,u}|,d_{k,u}\}$ elements from $T'_{k,u}$, and update $V^t_i=V^t_i\cup\{v_{k,u}\}$;
  \ENDWHILE
 \ENDFOR
\end{algorithmic}
\end{algorithm}

 Accordingly, we propose Algorithm \ref{alg_sche1} to solve \textbf{P2}. For each connected component $G_i$, we denote the set of traversed (scheduled) user nodes as $V^t_i$, and the rest untraversed user nodes are scheduled in a serial manner. Each time we select the node with maximum number of connected nodes in $V^t_i$ to schedule (line 5) till all nodes in one connected component are traversed. According to constraint (\ref{equ_P2}c), the set of available time slots $T'_{k,u}$ (line 6) is given by
\begin{align}\label{equ_available}
 T'_{k,u}=\left(\bigcup_{v_{k',u'}\in V^t_i,e_{(k,u),(k',u')}\in E}T_{k',u'}\right)^c,
\end{align}
 where $(\cdot)^c$ represents the complementary set.

 Next, the optimality of Algorithm \ref{alg_sche1} is proved in Theorem \ref{theo}. It can be seen that two conditions are required in the theorem. Firstly, the requirement $d_{k,u}$ should be well-designed such that the lower bound derived in Section \ref{sec_NIFSche_bound} is zero, which could be ensured in Section \ref{sec_SchePower}. Secondly, there is no empty cycle with $4$ or more nodes. Actually, we will show in Section \ref{sec_NIFSche_illu} that the empty cycles with $4$ or more nodes is very rare in $G$. It can be seen that the above two conditions can be generally guaranteed. Hence, Theorem \ref{theo} ensures that the optimal solution $n_0=0$ can be derived via Algorithm \ref{alg_sche1}, indicating that Algorithm \ref{alg_sche1} is near-optimal to solve \textbf{P2}.

\begin{theorem}\label{theo}
 If there is no clique node set $\{v_{k_q,u_q}\}_q$ such that $\sum_{q}d_{k_q,u_q}>N$ and no empty cycle with $4$ or more nodes, we can achieve $n_0=0$ by Algorithm \ref{alg_sche1}, i.e. $|T_{k,u}|=d_{k,u},\forall k,u$.
\end{theorem}
\begin{proof}
 See \cite{Proofs}.
\end{proof}

\subsection{Proposed Scheduling Algorithm}\label{sec_NIFSche_sche2}
 Back to problem \textbf{P1}, we need to consider the limited number $N_{RF}$ of RF chains. Based on Algorithm \ref{alg_sche1}, the rest problem is how to arrange $T_{k,u}$ in the $N_{RF}$ RF chains of each BS. Therefore, we extend Algorithm \ref{alg_sche1} to Algorithm \ref{alg_sche2} such that $T_{k,u}$ is selected based on the priority of each available time slot. The near-optimal property of Algorithm \ref{alg_sche1} is inherited by Algorithm \ref{alg_sche2}, so that the optimal solution $n_0=0$ can still be derived if $d_{k,u}$ is designed to satisfy the condition of Theorem \ref{theo}, which is verified by our simulation results in Section \ref{sec_NIFSche_illu}. Furthermore, for the scenario with unequal number of users in each cell, we can traverse $G$ and obtain the time slots allocated to each user in the same way, so Algorithm 3 can be directly adopted to this scenario.

 Additionally, although the optimization variables are $\{\bm{S}^k\}_{k=1}^{K}$ in \textbf{P1}, we still only need to solve $T_{k,u}$ to derive $\{\bm{S}^k\}_{k=1}^{K}$. Note that the order of the elements in one column of $\bm{S}^k$ does not influence \textbf{P1}. Therefore, once a user is scheduled at time slot $n$, the user can be served by any unoccupied RF chain. As a result, we only need to determine $T_{k,u}$ for each user.

\begin{algorithm}
\caption{Proposed NIF Space-time User Scheduling Algorithm}
\label{alg_sche2}
\begin{algorithmic}[1]
 \FOR{each connected component $G_i=\{V_i,E_i\}$ in $G$}
  \STATE Select one $v_{k_0,u_0}\in V_i$, and derive set $T'_{k_0,u_0}$ of the available time slots to schedule $v_{k,u}$.
  \STATE Compute the priority of the time slots in $T'_{k_0,u_0}$, and select $T_{k_0,u_0}$ accordingly;
  \WHILE{$V^t_i\ne V_i$}
   \STATE Search for the user nodes in $V_i\setminus V^t_i$ connected with $V^t_i$, from which select the user node $v_{k,u}$ with maximum number of connected nodes in $V^t_i$;
   \STATE Derive set $T'_{k,u}$ of the available time slot to schedule $v_{k,u}$, and update $V^t_i=V^t_i\cup\{v_{k,u}\}$;
   \STATE Compute the priority of the time slots in $T'_{k,u}$, and select $T_{k,u}$ accordingly;
  \ENDWHILE
 \ENDFOR
\end{algorithmic}
\end{algorithm}

\begin{figure}[tp]
 \begin{center}
 \includegraphics[width=0.95\columnwidth]{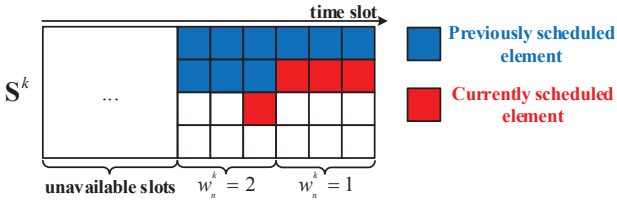}
 \end{center}
 \vspace{-5mm}
 \caption{An example of scheduling a user according to the priority of each available time slot. The time slots with low $w^k_n(=1)$ are assigned to the user firstly, and then the rest slots with $w^k_n=2$ can be assigned.}
  \vspace{-5mm}
 \label{fig_priority}
\end{figure}

 Then, we elaborate on how to determine the priority of the available time slots. Since the number of RF chains is limited, the major problem is to avoid more than $N_{RF}$ users in one cell scheduled at one time slot, otherwise there would be at least one user cannot be scheduled. Therefore, we define $w^k_n$ as
\begin{align}
 w^k_n=N_{RF}-\sum_{r}\delta(s^k_{r,n},0),
\end{align}
 indicating that $w^k_n$ RF chains of the $k$th BS have been occupied at time slot $n$. If $w^k_n=N_{RF}$, we cannot schedule any more user in cell $k$ at time slot $n$, and the time slot $n$ becomes an unavailable slot for the users in cell $k$, which should be taken into account in the computation of $T'_{k,u}$ (line 2, 6). To avoid $w^k_n$ reaching $N_{RF}$ quickly, in the selection of $T_{k,u}$, the time slots with low $w^k_n$ are selected firstly with high priority, as illustrated in Fig.~\ref{fig_priority}. Therefore, the priority of available time slots (line 3, 7) is given by the ascending order of $w^k_n$ of the time slots. Then, we select $\text{min}\{|T'_{k,u}|,d_{k,u}\}$ time slots from $T'_{k,u}$ according to the priority to derive $T_{k,u}$.

\subsection{Illustrative Results and Discussions}\label{sec_NIFSche_illu}
\begin{figure}[tp]
 \begin{center}
 \includegraphics[width=0.7\columnwidth]{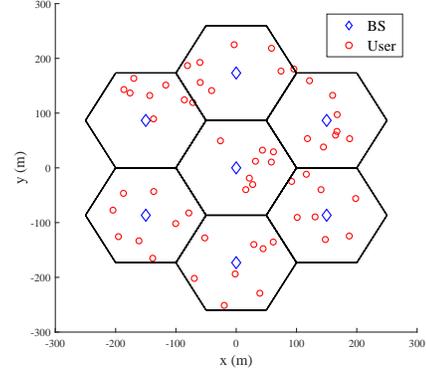}
 \end{center}
 \vspace{-5mm}
 \caption{One realization of simulation scenario with $K=7$, $U=8$.}
 \label{fig_networkSim}
 \vspace{-5mm}
\end{figure}

 Monte Carlo simulations are conducted to illustrate the performance of our proposed space-time scheduling Algorithm \ref{alg_sche2}, in terms of the unfulfilled user requirements on resource elements. We consider a network with $K=7$ regular hexagon cells with radius of $100$m, as shown in Fig.~\ref{fig_networkSim}, where $U=8,12,16$ users are randomly located in each cell with uniform distribution. The path loss $L^k_{k',u}$ between the $k$th BS and the $u$th user in cell $k'$ is model according to the 3GPP urban macro model \cite{3GPP}. Specifically, the path losses result from LoS and NLoS channels are given by
\begin{align}
 L^{(\text{LoS})}(\text{dB})&=32.4+20\text{log}_{10}(d)+20\text{log}_{10}(f_c),\\
 L^{(\text{NLoS})}(\text{dB})&=13.54+39.08\text{log}_{10}(d)+20\text{log}_{10}(f_c),
\end{align}
 where $f_c=28$GHz represents the carrier frequency, and $d$ is the distance between BS and user. Since the mmWave transmission is commonly LoS-dominant, the channel between a user and its serving BS is considered as LoS channel. Meanwhile, the channel between a user and a remote BS has probability $\text{P}^{(\text{LoS})}$ to be LoS and $(1-\text{P}^{(\text{LoS})})$ to be NLoS, expressed as
\begin{align}
 \text{P}^{(\text{LoS})}=\frac{18}{d}+\text{exp}\left(-\frac{d}{63}\left(1-\frac{18}{d}\right)\right),
\end{align}
 implying that there might be blockage or reflection in the channel. Besides, each BS uses $2N_t=32$ beams to cover the users from all directions, as shown in Fig.~\ref{fig_beam}, and the number of RF chains equipped by each BS is $N_{RF}=2,3,4$. One scheduling period consists of $N=16$ time slots.

\begin{figure}[tp]
 \begin{center}
 \includegraphics[width=0.7\columnwidth]{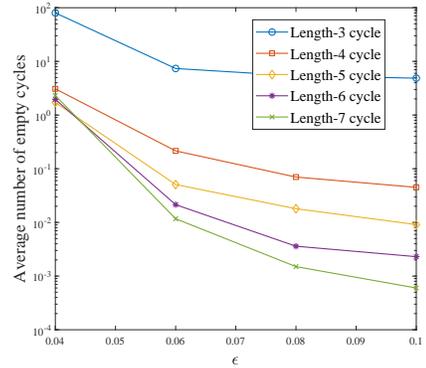}
 \end{center}
 \vspace{-5mm}
 \caption{The average number of empty cycles in $G$ under various $\epsilon$, with $U=8$.}
 \vspace{-5mm}
 \label{fig_emptyCycle}
\end{figure}

 According to Theorem \ref{theo}, the number of empty cycles with $4$ or more nodes in $G$ should be as less as possible, which requires a relatively large threshold $\epsilon$ to ensure the sparsity of $G$. However, a large $\epsilon$ may lead to too many ignored interferences. As a result, we aim to select a relatively small value of $\epsilon$ from the possible values which lead to few empty cycles with $4$ or more nodes. The number of empty cycles in $G$ under various $\epsilon$ is illustrated in Fig.~\ref{fig_emptyCycle}.
 It can be seen that the numbers of the empty cycles with $4$ or more nodes reduce to a very low level when $\epsilon\geq0.06$. Especially, for $\epsilon=0.08$, the average number of length-$4$ empty cycles is less than $10^{-1}$, indicating that the empty cycles with $4$ or more nodes are very rare in $G$. Hence, in the following simulations, we set $\epsilon=0.08$ as a tradeoff between the tolerance on interference and the sparsity of interference graph.

\begin{figure*}[tp]
	\begin{center}
		\includegraphics[width=1.5\columnwidth]{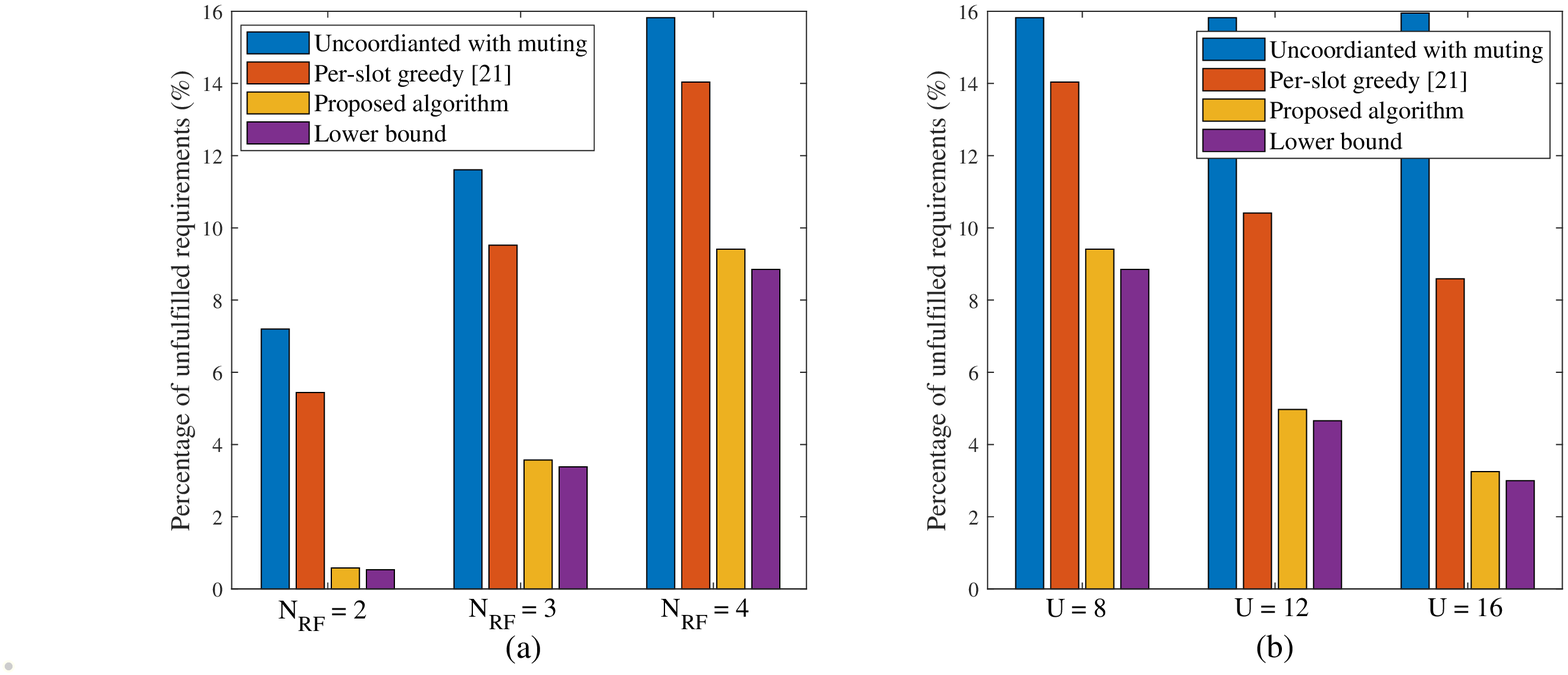}
	\end{center}
	\vspace{-5mm}
	\caption{The percentage of unfulfilled requirements when $d_{k,u}$ is randomly generated. (a) $N_{RF}=2,3,4$, $U=8$; (b) $N_{RF}=4$, $U=8,12,16$.}
	\label{fig_NcRand}
	\vspace{-5mm}
\end{figure*}

\begin{figure*}[tp]
	\begin{center}
		\includegraphics[width=1.5\columnwidth]{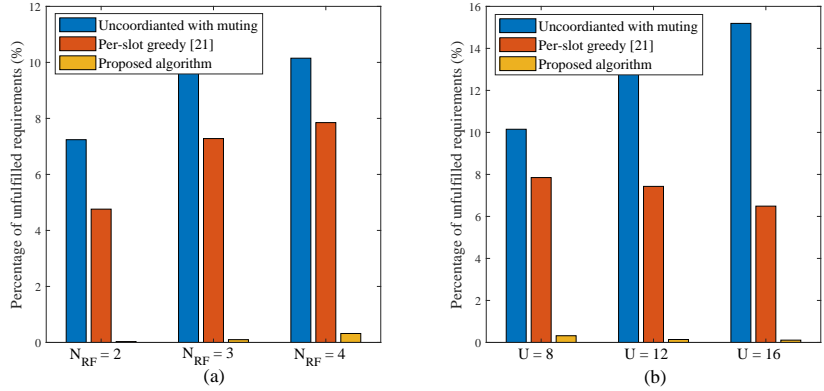}
	\end{center}
	\vspace{-5mm}
	\caption{The percentage of unfulfilled requirements when $d_{k,u}$ is randomly generated ensuring that the lower bound is $0$. (a) $N_{RF}=2,3,4$, $U=8$; (b) $N_{RF}=4$, $U=8,12,16$.}
	\vspace{-5mm}
	\label{fig_NcZeroBound}
\end{figure*}

 Next, we evaluate the performance of Algorithm \ref{alg_sche2}. We compare our proposed algorithm with the per-slot greedy scheduling \cite{low5}, the scheduling without cell coordination, and the lower bound derived in Section \ref{sec_NIFSche_bound}. In the per-slot greedy algorithm, the scheduling is performed in a per-slot manner, and the scheduled user in each time slot is selected greedily from $G$. In the uncoordinated scheduling, each cell schedules its local users to avoid intra-cell interferences, while the inter-cell interferences are avoided by muting resource elements.

 The percentage of unfulfilled requirements over all resource elements, i.e. $\frac{n_0}{KN_{RF}N}\times100\%$, is simulated under different $U$ and $N_{RF}$. In Fig.~\ref{fig_NcRand}, we generate the requirement $0\leq d_{k,u}\leq N$ of each user randomly but satisfying (\ref{equ_dPerCell}). It can be seen that the proposed algorithm outperforms the other two schemes under various parameters, and is close to the lower bound as well, which indicates that Algorithm \ref{alg_sche2} is near-optimal and the lower bound is tight. Besides, in Fig.~\ref{fig_NcRand}(a), the percentage of unfulfilled requirements increases with $N_{RF}$ because more beams are used simultaneously, leading to larger probability of interference. Meanwhile, in Fig.~\ref{fig_NcRand}(b), each user shares a smaller $d_{k,u}$ with the growth of $U$, so it is more difficult to form a target clique in the computation of the lower bound (Algorithm \ref{alg_bound}), and thus the unfulfilled requirements decrease.

 Furthermore, we randomly generate the $d_{k,u}$ of each user ensuring that the lower bound is $0$. The corresponding performance is shown in Fig.~\ref{fig_NcZeroBound}. The number of unfulfilled requirements in Algorithm \ref{alg_sche2} is nearly $0$ under various parameters, while there are $5\sim8\%$ and $7\sim15\%$ unfulfilled requirements for the greedy schedule and the uncoordinated schedule, respectively.

 More importantly, Fig.~\ref{fig_NcZeroBound} implies that we can almost ensure that there is no unfulfilled requirement via Algorithm \ref{alg_sche2}, if $d_{k,u}$ is well-designed so that the lower bound is $0$. Therefore, to achieve zero unfulfilled requirement, no clique with sum requirements exceeding $N$ is not only a necessary condition, but also a near-sufficient one, which is an important result for the design of $d_{k,u}$ in Section \ref{sec_SchePower}.

\section{Joint Space-Time User Scheduling and Power Allocation}\label{sec_SchePower}

 In this section, we study the joint NIF user scheduling and power allocation problem to minimize the total transmit power. Instead of assuming that each user has a requirement on resource element, we consider a more general user requirement on data rate. When the rate requirement of one user is fixed, assigning more resource elements to the user leads to lower transmit power. Based on our proposed NIF scheduling, an energy-efficient joint scheduling and power allocation scheme can be derived.

\subsection{CSI Acquisition}\label{sec_SchePower_CSI}
 The measurement of an inter-cell interference link is resource-costing in practical implementation, making the CSI acquisition a challenging task\cite{low2}. Therefore, it is hard to acquire the CSI between every BS and user in the network, even if only partial or large-scale CSI is required \cite{mmWave8}. Consequently, we may only have limited information on a limited number of inter-cell interference links, which obstructs the optimization of network.

 However, in NIF scheduling, the total interference power at the user node is low because all strong interferences are avoided. As a result, we can assume a maximum overall interference power $I_{k,u}$ experienced by each user in each scheduling period. To acquire $I_{k,u}$, one user only needs to measure the total received interference (plus noise) power on the local zero-power RS, where the coordinated measurement between BSs is not required. Therefore, we denote the local channel gain and the interference-plus-noise power of each user as $\alpha_{k,u}\triangleq L^k_{k,u}G^{k,u}_{k,u}$ and $\tilde{I}_{k,u}\triangleq I_{k,u}+p_n$, respectively, so the SINR in (\ref{equ_SINR}) for one resource element can be simplified as
\begin{align}\label{equ_SINRSimple}
 \rho^{k}_{r,n}=\frac{\alpha_{k,s^k_{r,n}}}{\tilde{I}_{k,s^k_{r,n}}}p^k_{r,n}.
\end{align}
 After each scheduling period, every user feeds back the maximum measured interference-plus-noise power $\tilde{I}_{k,u}$, and the channel gain $\alpha_{k,u}$ to its local BS, which form the CSI for the next scheduling period. Furthermore, (\ref{equ_SINRSimple}) is also applicable to the scenario with Tx digital precoding. In this case, the $\alpha_{k,u}$ is the equivalent channel gain after digital precoding, and $\tilde{I}_{k,u}$ still represents the maximum measured interference-plus-noise power, while $\tilde{I}_{k,u}$ would be further reduced since the intra-cell interference can be cancelled by zero-forcing (ZF) digital precoding. It can be seen that the users do not need to measure the CSI of inter-cell interference links during the scheduling in our framework. Meanwhile, according to Section \ref{sec_system_graph}, the overhead induced by the establishment of $G$ is also acceptable.

 By replacing the sum power of weak interferences and noise by a constant value, the joint user scheduling and power allocation problem can be converted into an integer convex optimization and solved near-optimally, as demonstrated in Section \ref{sec_SchePower_problem}. Besides, it could be validated that the performance loss induced by this simplification is negligible \cite{Proofs}.

\subsection{Problem Formulation and Transformation}\label{sec_SchePower_problem}
 Combining (\ref{equ_averRate}), (\ref{equ_SINRSimple}) and \textbf{P1}, the joint NIF user scheduling and power allocation problem \textbf{P3} to minimize the total transmit power is formulated as
\begin{align}\label{equ_P3}
 \textbf{P3:}~~&\min\limits_{\{\bm{S}^k,\bm{P}^k\}_{k=1}^{K}}~\sum_{(k,r,n)}p^k_{r,n}\\ \nonumber \tag{\ref{equ_P3}a}
 \text{s.t.}~~&~s^{k}_{r,n}\in\{0,1,\cdots,U\},~\forall k,r,n\\ \nonumber \tag{\ref{equ_P3}b}
 &\delta(s^k_{r,n},s^{k}_{r',n})=\delta(r,r'),~\text{if}~s^k_{r,n}s^{k}_{r',n}\ne0,~\forall k,r,r',n\\ \nonumber \tag{\ref{equ_P3}c}
 &e_{(k,s^k_{r,n}),(k',s^{k'}_{r',n})}\notin E,~\text{if}~s^k_{r,n}s^{k'}_{r',n}\ne0,~\forall k,r,k',r',n\\ \nonumber \tag{\ref{equ_P3}d}
 &0\leq p^k_{r,n}\leq P_{max},~\forall k,r,u\\ \nonumber \tag{\ref{equ_P3}e}
 &\frac{1}{N}\sum_{(r,n),s^k_{r,n}=u}W\text{log}_2(1+\beta_{k,s^k_{r,n}}p^k_{r,n})\geq\gamma_{k,u}~\forall k,u,
\end{align}
 where (\ref{equ_P3}a), (\ref{equ_P3}b), (\ref{equ_P3}c) are inherited from \textbf{P1} ensuring NIF scheduling. (\ref{equ_P3}d) requires that the power allocated to each resource element does not exceed the maximum output power $P_{max}$ of a RF chain.
 Based on (\ref{equ_SINRSimple}), we define $\beta_{k,s^k_{r,n}}\triangleq\frac{\alpha_{k,s^k_{r,n}}}{\tilde{I}_{k,s^k_{r,n}}}$ (namely $\beta_{k,u}\triangleq\frac{\alpha_{k,u}}{\tilde{I}_{k,u}}$), and thus derive the constraints on user rate requirement in (\ref{equ_P3}e). Since the maximum interference-plus-noise power $\tilde{I}_{k,u}$ is used to replace the interference-plus-noise power for each user, (\ref{equ_P3}e) ensures that our derived user scheduling and power allocation can fulfill the rate requirement of each user. Next, we transform the constraints on rate requirement to the constraints on $d_{k,u}$.

 Obviously, \textbf{P3} is a mix-integer optimization problem which is difficult to solve directly, but we can simplify \textbf{P3} via the concavity of (\ref{equ_P3}e) with respect to $p^k_{r,n}$. Note that the power allocations to different users are decoupled in (\ref{equ_P3}e). So, we let the scheduling $\{\bm{S}^k\}_{k=1}^{K}$ be fixed, and consider the power allocation to one user node $v_{k,u}$. According to the concavity of (\ref{equ_P3}e), the powers allocated to every resource element of $v_{k,u}$ should be the same, otherwise we can find a new set of $\{p^k_{r,n}\}_{s^k_{r,n}=u}$ with less $\sum_{s^k_{r,n}=u}p^k_{r,n}$ and still satisfying (\ref{equ_P3}e), so we have
\begin{align}\label{equ_pEqual}
 p^k_{r,n}=p^k_u,~~\text{for}~s^k_{r,n}=u,
\end{align}
 where $p^k_u$ denotes the power allocated to each resource element assigned to $v_{k,u}$. Then, according to (\ref{equ_dConstraint}) and (\ref{equ_pEqual}), (\ref{equ_P3}e) can be rewritten as
\begin{align}\label{equ_rateEqual}
 \frac{\hat{d}_{k,u}}{N}W\text{log}_2(1+\beta_{k,u}p^k_{u})\geq\gamma_{k,u}.
\end{align}
 Note that the left-hand side of (\ref{equ_rateEqual}) is monotonically increasing with $p^k_{u}$, so the right-hand and left-hand sides of (\ref{equ_rateEqual}) should be equal to minimize the total transmit power, leading to
\begin{align}\label{equ_pEqual2}
 p^k_u=\frac{1}{\beta_{k,u}}\left(2^{\frac{\gamma_{k,u}N}{\hat{d}_{k,u}W}}-1\right).
\end{align}
 Furthermore, according to (\ref{equ_pEqual2}) and $p^k_u\leq P_{max}$, (\ref{equ_P3}d) can be converted into
\begin{align}\label{equ_dConstraint2}
 \hat{d}_{k,u}\geq\left\lceil\frac{\gamma_{k,u}N}{W\text{log}_2(\beta_{k,u}P_{max}+1)}\right\rceil,
\end{align}
 where $\lceil x\rceil$ denotes the minimum integer not less than $x$.
 Then, \textbf{P3} can be transformed to
\begin{align}\label{equ_P4}
 \textbf{P4:}~~&\min\limits_{\{\bm{S}^k\}_{k=1}^{K}}~\sum_{(k,u)}\frac{\hat{d}_{k,u}}{\beta_{k,u}}\left(2^{\frac{\gamma_{k,u}N}{\hat{d}_{k,u}W}}-1\right)\\ \nonumber \tag{\ref{equ_P4}a}
 \text{s.t.}~~&~s^{k}_{r,n}\in\{0,1,\cdots,U\},~\forall k,r,n\\ \nonumber \tag{\ref{equ_P4}b}
 &\delta(s^k_{r,n},s^{k}_{r',n})=\delta(r,r'),~\text{if}~s^k_{r,n}s^{k}_{r',n}\ne0,~\forall k,r,r',n\\ \nonumber \tag{\ref{equ_P4}c}
 &e_{(k,s^k_{r,n}),(k',s^{k'}_{r',n})}\notin E,~\text{if}~s^k_{r,n}s^{k'}_{r',n}\ne0,~\forall k,r,k',r',n\\ \nonumber \tag{\ref{equ_P4}d}
 &\hat{d}_{k,u}=\sum_{(r,n)}\delta(s^{k}_{r,n},u),~\forall k,u\\ \nonumber \tag{\ref{equ_P4}e}
 &\left\lceil\frac{\gamma_{k,u}N}{W\text{log}_2(\beta_{k,u}P_{max}+1)}\right\rceil\leq \hat{d}_{k,u}\leq N,~~\forall k,u,
\end{align}
 It can be seen that the variables to be optimized only include $\{\bm{S}^k\}_{k=1}^{K}$ after the transformation.

 Next, we can utilize the results derived from Section \ref{sec_NIFSche} to further simplify \textbf{P4}. Since the objective function (\ref{equ_P4}) is only determined by $\hat{d}_{k,u}$, we consider problem \textbf{P5} given by
\begin{align}\label{equ_P5}
 \textbf{P5:}~~&\min\limits_{\{d_{k,u}\}_{k,u}}~\sum_{(k,u)}\frac{d_{k,u}}{\beta_{k,u}}\left(2^{\frac{\gamma_{k,u}N}{d_{k,u}W}}-1\right)\\ \nonumber \tag{\ref{equ_P5}a}
 &\left\lceil\frac{\gamma_{k,u}N}{W\text{log}_2(\beta_{k,u}P_{max}+1)}\right\rceil\leq d_{k,u}\leq N,d_{k,u}\in\mathbb{N},\forall k,u\\ \nonumber \tag{\ref{equ_P5}b}
 &\sum_{q}d_{k_q,u_q}\leq N,~\text{for any clique with node set}~\{v_{k_q,u_q}\}_q\\ \nonumber \tag{\ref{equ_P5}c}
 &\sum_{u}d_{k,u}\leq N_{RF}N,~\forall k.
\end{align}
 We denote the optimal value of \textbf{P4} and \textbf{P5} as $p^*_4$ and $p^*_5$, respectively. For any feasible $\{\bm{S}^k\}_{k=1}^{K}$ and the corresponding $\hat{d}_{k,u}$ in \textbf{P4}, by letting $d_{k,u}=\hat{d}_{k,u}$, we derive a feasible set of $\{d_{k,u}\}_{k,u}$ for \textbf{P5}, where (\ref{equ_P5}b) holds according to Proposition \ref{prop_clique}, and (\ref{equ_P5}a), (\ref{equ_P5}c) hold naturally. Hence, we have $p^*_5\leq p^*_4$. Conversely, for any feasible $\{d_{k,u}\}_{k,u}$ in \textbf{P5}, according to the results in Fig.~\ref{fig_NcZeroBound}, we can almost ensure that a scheduling $\{\bm{S}^k\}_{k=1}^{K}$ can be found by Algorithm \ref{alg_sche2} such that $\hat{d}_{k,u}=d_{k,u}$ and (\ref{equ_P4}a)-(\ref{equ_P4}d) hold, while (\ref{equ_P4}e) holds naturally.
 Therefore, one feasible solution to \textbf{P5} is generally also feasible for \textbf{P4}, but this feasibility is not strictly guaranteed, leading to $p^*_5\approx p^*_4$. Given that the optimal solution $p^*_3$ to \textbf{P3} equals to $p^*_4$, the gap between $p^*_3$ and $p^*_5$ is small.

\subsection{Proposed Scheme}\label{sec_SchePower_prop}
 Firstly, we derive the user requirements $\{d_{k,u}\}_{k,u}$ on resource elements by solving \textbf{P5}. It can be seen that the objective function (\ref{equ_P5}) is convex with respect to $d_{k,u}$ when $d_{k,u}>0$. Hence, \textbf{P5} is an integer convex optimization problem. By relaxing the integer constraint on $d_{k,u}$, we can derive the near-optimal $d_{k,u}$ for \textbf{P5}, which can be further converted into the near-optimal $\{\bm{S}^k\}_{k=1}^{K}, \{\bm{P}^k\}_{k=1}^{K}$ for \textbf{P3}.
 After the relaxing of integer constraint, the global minimum solution $\tilde{p}^*_5$ to \textbf{P5} can be readily found via convex optimization algorithms \cite{convex}. Clearly, we have $\tilde{p}^*_5\leq p^*_5$, so $\tilde{p}^*_5$ can be considered as a lower bound of $\textbf{P5}$. Furthermore, since $p^*_5\leq p^*_4$, $\tilde{p}^*_5$ is also the lower bound of \textbf{P4} and \textbf{P3}.

 We denote the continuous requirements derived by convex optimization as $\{\tilde{d}_{k,u}\}_{k,u}$, and the near-optimal $d_{k,u}$ can be given by
\begin{align}\label{equ_round}
 d_{k,u}=\lceil\tilde{d}_{k,u}\rfloor,~~\forall k,u,
\end{align}
 where $\lceil x\rfloor$ represents the closest integer to $x$. However, the $\{d_{k,u}\}_{k,u}$ derived by (\ref{equ_round}) may violate the constraints (\ref{equ_P5}b), (\ref{equ_P5}c), so a further adjustment is required.

 Accordingly, Algorithm \ref{alg_alloc} is proposed to solve $d_{k,u}$. After the integer approximation in (\ref{equ_round}), we satisfy the violated constraints by reducing $d_{k,u}$. Specifically, we repeat selecting one $d_{k_0,u_0}$ from the violated (\ref{equ_P5}b) or (\ref{equ_P5}c), and reducing $d_{k_0,u_0}$ by $1$, until no (\ref{equ_P5}b) or (\ref{equ_P5}c) is violated. Note that the objective function (\ref{equ_P5}) is monotonically decreasing with the growth of $d_{k,u}$, so the reduction of $d_{k_0,u_0}$ leads to the increase of total transmit power. Therefore, in the selection of $d_{k_0,u_0}$, we always search for the $d_{k_0,u_0}$ whose reduction causes the minimum increment in the total transmit power $p_5$ given by (\ref{equ_P5}), and meanwhile does not violate (\ref{equ_P5}a).

\begin{algorithm}
\caption{Resource Allocation Algorithm}
\label{alg_alloc}
\begin{algorithmic}[1]
 \STATE Relax the integer constraint in \textbf{P5}, and derive $\tilde{d}_{k,u}$ via convex optimization algorithm;
 \STATE Set $d_{k,u}=\lceil\tilde{d}_{k,u}\rfloor$, and compute the current value $p_5$ of (\ref{equ_P5});
 \WHILE{there is one violated (\ref{equ_P5}b) or (\ref{equ_P5}c)}
  \STATE Select the $d_{k_0,u_0}$ from the violated (\ref{equ_P5}b) or (\ref{equ_P5}c) to be reduced by $1$, which causes the minimum increment in $p_5$ without violating (\ref{equ_P5}a);
  \STATE Set $d_{k_0,u_0}=d_{k_0,u_0}-1$, and update $p_5$;
 \ENDWHILE
\end{algorithmic}
\end{algorithm}

 After the derivation of $d_{k,u}$, we can obtain the scheduling $\{\bm{S}^k\}_{k=1}^{K}$ via Algorithm \ref{alg_sche2}. According to Section \ref{sec_NIFSche_illu}, although we can almost ensure zero unfulfilled requirement, there might still be few unfulfilled requirements, leading to unassigned resource elements in $\{\bm{S}^k\}_{k=1}^{K}$. Therefore, we propose Algorithm \ref{alg_palloc} which assigns the unassigned resource elements to users so that no resource element is wasted, and then derives the power allocation $\{\bm{P}^k\}_{k=1}^{K}$.  Similar to Algorithm \ref{alg_alloc}, since (\ref{equ_P4}) decreases with the growth of $d_{k,u}$, we select the user $v_{k,u_0}$ for an unassigned resource element $s^k_{r,n}=0$ under NIF condition, which causes the maximum decrement in the total transmit power $p_4$ given by (\ref{equ_P4}). After the scheduling $\{\bm{S}^k\}_{k=1}^{K}$ is determined, the power allocation could be derived according to (\ref{equ_pEqual}) and (\ref{equ_pEqual2}).

\begin{algorithm}
\caption{Scheduling Adjustment and Power Allocation Algorithm}
\label{alg_palloc}
\begin{algorithmic}[1]
 \STATE Compute the current value $p_4$ of (\ref{equ_P4});
 \WHILE{there is one $s^k_{r,n}=0$}
  \STATE Select one user node $v_{k,u_0}$ from the nodes not connected with the scheduled nodes in time slot $n$, which causes the maximum decrement in $p_4$;
  \STATE Set $s^k_{r,n}=u_0$, and update $p_4$;
 \ENDWHILE
 \STATE Derive the power allocation $\{\bm{P}^k\}_{k=1}^{K}$ according to (\ref{equ_pEqual}) and (\ref{equ_pEqual2});
\end{algorithmic}
\end{algorithm}

\begin{figure}[tp]
 \begin{center}
 \includegraphics[width=0.6\columnwidth]{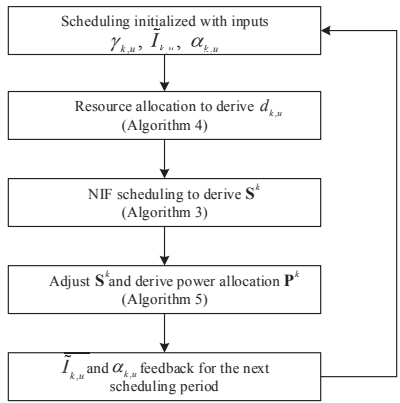}
 \end{center}
 \vspace{-5mm}
 \caption{The procedure of our proposed joint NIF user scheduling and power allocation scheme for \textbf{P3}.}
 \label{fig_wholeScheme}
 \vspace{-5mm}
\end{figure}

 As a summary of Section IV A-C, the entire procedure of our proposed joint NIF user scheduling and power allocation scheme is illustrated in Fig.~\ref{fig_wholeScheme}. For each scheduling period, the scheduling is initialized by the rate requirement $\gamma_{k,u}$, and the CSI $\tilde{I}_{k,u}$, $\alpha_{k,u}$. By solving \textbf{P5} via Algorithm \ref{alg_alloc}, the user requirements on resource elements $d_{k,u}$ could be derived, where the derived $d_{k,u}$ is near-optimal because of the convexity of \textbf{P5}. Given $d_{k,u}$, the NIF scheduling $\bm{S}^k$ could be derived using Algorithm \ref{alg_sche2} with only few requirements $d_{k,u}$ unfulfilled. Finally, by Algorithm \ref{alg_palloc}, the unassigned resource elements are assigned to users, and the power allocation on each resource element could be computed. The measured $\tilde{I}_{k,u}$, $\alpha_{k,u}$ in the current scheduling period are fed back as the CSI for the next scheduling period.

 Additionally, for the scenario with unequal number of users in each cell, we only need to set different number of variables $d_{k,u}$ for each cell, which does not affect the design of Algorithm \ref{alg_sche2}-\ref{alg_palloc}. Therefore, the extension of our proposed joint user scheduling and power allocation scheme to unequal number of users in each cell is validated.

\subsection{Feasibility Issue}\label{sec_SchePower_feas}
 Since the power allocated to each resource element is upper bounded by $P_{max}$, the data rate of one user cannot increase unlimitedly. As a result, the joint scheduling and power allocation problem could be infeasible under certain rate requirements $\gamma_{k,u}$, e.g. a user with poor channel condition $\beta_{k,u}$ has high rate requirement. Towards this issue, our strategy is to fulfill the rate requirements in best effort if the problem is found infeasible.

 Specifically, before solving \textbf{P5}, the feasibility of \textbf{P5} could be checked by letting
\begin{align}
 d_{k,u}=\left\lceil\frac{\gamma_{k,u}N}{W\text{log}_2(\beta_{k,u}P_{max}+1)}\right\rceil,
\end{align}
 and judging whether (\ref{equ_P5}b), (\ref{equ_P5}c) are violated. If there is one violated (\ref{equ_P5}b) or (\ref{equ_P5}c), \textbf{P5} is infeasible. In this case, the constraint (\ref{equ_P5}a) is relaxed to $0\leq d_{k,u}\leq N,~d_{k,u}\in\mathbb{N}$, implying that the power constraint $P_{max}$ on each resource element is relaxed, which ensures that \textbf{P5} is feasible and $d_{k,u}$ could be solved via Algorithm \ref{alg_alloc}. Next, since the derived $0\leq d_{k,u}\leq N$ satisfies (\ref{equ_P5}c), Algorithm \ref{alg_sche2} is always feasible to output the scheduling $\bm{S}^k$. Finally, we retrieve the power constraint on each resource element and limit the output power allocation in Algorithm \ref{alg_palloc} by
\begin{align}
 p^k_{r,n}=\text{min}\{p^k_{r,n},P_{max}\},~~\forall k,r,n,
\end{align}
 which guarantees that the power allocated to each resource element does not exceed $P_{max}$. In this way, we derive the user scheduling and power allocation scheme in best effort.

\section{Simulation Results}\label{sec_sim}
 In this section, the performance of our proposed joint NIF user scheduling and power allocation scheme is evaluated in comparison with the existing method via Monte Carlo simulation. We adopt the same network and path loss model as Section \ref{sec_NIFSche_illu}. The rate requirement $\gamma_{k,u}$ of each user is randomly generated by $\frac{WN_{RF}X}{U}$ with uniformly distributed random variable $X\sim U(1,4)$, corresponding to the average spectral efficiency of one single RF chain varying from $1$bit/s/Hz to $4$bit/s/Hz randomly. Our simulation parameters are summarized in Table I.

\begin{table}[tp]
\centering
\caption{Simulation Parameters}
\vspace*{-3mm}
\begin{tabular}{cc}
 \hline
 {Parameter} & {Value}\\
 \hline
 {Carrier frequency $f_c$} & {$28$GHz} \\
 {Bandwidth $W$} & {$250$MHz} \\
 {Cell number $K$} & {$7$} \\
 {Cell radius $l$} & {$100$m} \\
 {RF chain number $N_{RF}$} & {$4$} \\
 {User number $U$ per cell} & {$8$} \\
 {Beam number $2N_t$ per cell} & {$32$} \\
 {Side-lobe beam power gain $g_{min}$} & {$-6$dB} \\
 {Receiver noise figure} & {$6$dB} \\
 {Noise power spectrum density} & {$-174$dBm/Hz} \\
 {Scheduling period number} & {$20$} \\
 {Time slot number $N$ per period} & {$16$} \\
 {Transmit power $P_{max}$} & {$24$-$30$dBm} \\
 {Interference threshold $\epsilon$} & {$0.08$} \\
 \hline
\end{tabular}
\vspace*{-5mm}
\end{table}


 For comparison, the performance of the joint user scheduling and power allocation scheme proposed in \cite{mmWave1} is evaluated in our simulation. The scheduling scheme in \cite{mmWave1} is based on the ISs in $G$, where the users are grouped into several ISs and the scheduling in each time slot is given by one of the ISs. Then, the proportion of the time slots allocated to each IS and the corresponding power allocation are derived according to the rate requirements of users. Additionally, the original scheme \cite{mmWave1} may results in many unassigned resource elements, so its output is enhanced by our Algorithm \ref{alg_palloc} in the simulations.

\subsection{Single Scheduling Period}\label{sec_sim_single}
 Firstly, we study the performance of our proposed scheme on solving the joint scheduling and power allocation problem \textbf{P3}, corresponding to the performance in single scheduling period. Since the initial $\tilde{I}_{k,u}$ of the $1$st scheduling period is unknown, we set the initial $I_{k,u}$ coarsely as
\begin{align}\label{equ_initialI}
 \tilde{I}_{k,u}=\eta\sum_{k'}N_{RF}P_0L^{k'}_{k,u}g_{min}+p_n,
\end{align}
 where $\eta=0.2$ is a parameter of the strength of initial $\tilde{I}_{k,u}$, and $P_0=24$dBm. Although the initial $\tilde{I}_{k,u}$ given by (\ref{equ_initialI}) could be inaccurate because the real $\tilde{I}_{k,u}$ has not been measured, we will show in Section \ref{sec_sim_consecutive} that the network performances under different initial $\tilde{I}_{k,u}$ would converge to the same value after a few scheduling periods. In this subsection, we compare the performance of our scheme with the scheme in \cite{mmWave1} under the initial $\tilde{I}_{k,u}$ given by (\ref{equ_initialI}).

\begin{figure*}[tp]
 \begin{center}
 \includegraphics[width=1.5\columnwidth]{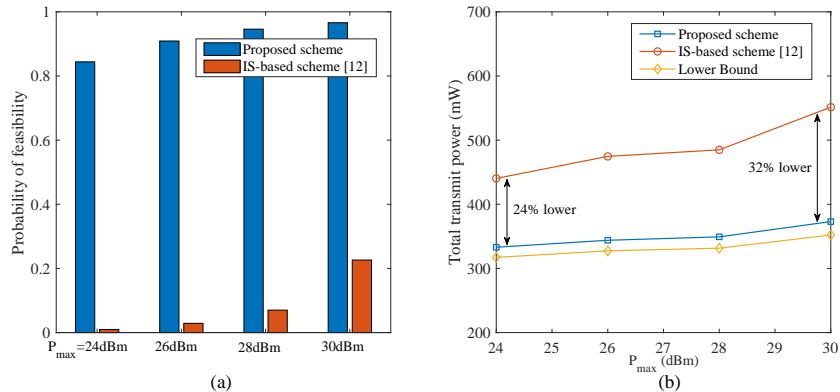}
 \end{center}
 \vspace{-5mm}
 \caption{Network performance comparison in single scheduling period. (a) Probability of feasibility; (b) Total transmit power.}
 \vspace{-5mm}
 \label{fig_single}
\end{figure*}

\begin{figure*}[tp]
	\begin{center}
		\includegraphics[width=1.5\columnwidth]{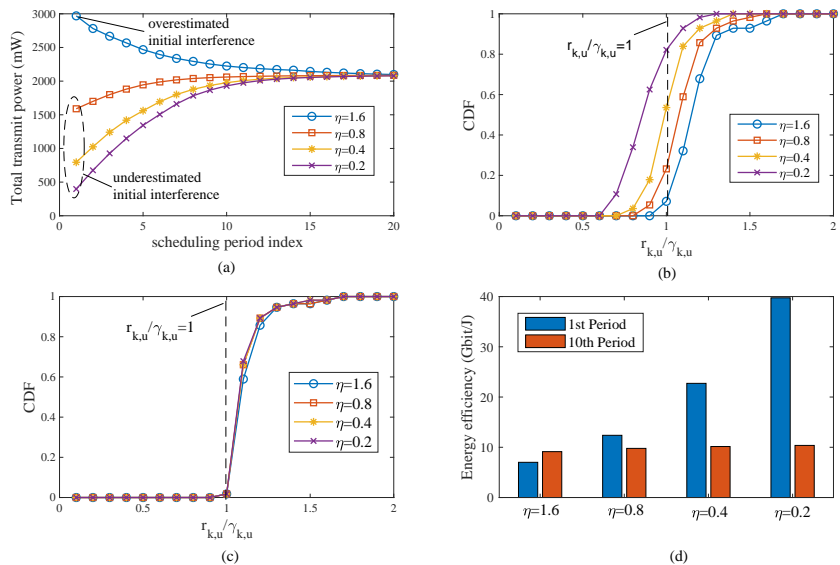}
	\end{center}
	\vspace{-5mm}
	\caption{Performance convergence in consecutive periods with different $\eta$ in one network realization. (a) Total transmit power; (b) CDF of $\frac{r_{k,u}}{\gamma_{k,u}}$ in the $1$st scheduling period; (c) CDF of $\frac{r_{k,u}}{\gamma_{k,u}}$ in the $10$th scheduling period; (d) Energy efficiency.}
	\vspace{-5mm}
	\label{fig_iFactor}
\end{figure*}

 In Fig.~\ref{fig_single}(a), the probabilities of the feasibility of solving \textbf{P3} via our proposed scheme and the IS-based scheme under random $\gamma_{k,u}$ are studied. Obviously, the proposed scheme outperforms the IS-based scheme significantly. Since the feasibility is mainly influenced by the maximum Tx power $P_{max}$, it is clear that our scheme could utilize the space-time resource elements more efficiently, preventing the power on single resource element from exceeding $P_{max}$.

 In Fig.~\ref{fig_single}(b), the total transmit power is investigated, where we only count the randomly given $\gamma_{k,u}$ which is feasible for both our scheme and the IS-based scheme. As mentioned in Section \ref{sec_SchePower_prop}, the optimal value $\tilde{p}^*_5$ of \textbf{P5} after relaxing the integer constraint is given as the lower bound of the total transmit power in \textbf{P3}. It can be seen that the total transmit power of our scheme is lower than \cite{mmWave1} by $24\%\sim32\%$. Meanwhile, our scheme can achieve a near-optimal performance which is close to the lower bound $\tilde{p}^*_5$ as shown in Fig.~\ref{fig_single}(b).

\subsection{Consecutive Scheduling Periods}\label{sec_sim_consecutive}

 Next, we study the network performance in consecutive scheduling periods. The $\tilde{I}_{k,u}$ and $\alpha_{k,u}$ would be updated after each scheduling period. In the following simulations, we allow \textbf{P3} to be infeasible under certain $\gamma_{k,u}$, where the best effort strategy introduced in Section \ref{sec_SchePower_feas} is adopted if \textbf{P3} is found infeasible. To evaluate the fulfillment of user rate requirements, the cumulative distribution function (CDF) of $\frac{r_{k,u}}{\gamma_{k,u}}$ for all users is studied. Clearly, if $\frac{r_{k,u}}{\gamma_{k,u}}\geq1$, the rate requirement $\gamma_{k,u}$ is fulfilled.

 The network performances with different $\eta$ in (\ref{equ_initialI}) under one fixed network realization are illustrated in Fig.~\ref{fig_iFactor}, where $P_{max}=24$dBm. As shown in Fig.~\ref{fig_iFactor}(a), for different $\eta$, the initial interference level $\tilde{I}_{k,u}$ could be overestimated or underestimated. However, the total transmit power for different $\eta$ all converge to the same value after around $10$ periods. In Fig.~\ref{fig_iFactor}(b), since the initial $\tilde{I}_{k,u}$ is inaccurate, the rate requirement of a considerable proportion of users cannot be fulfilled in the $1$st period, while the actual rate of some users may exceed the required rate obviously, leading to the waste of power. Nevertheless, in the $10$th period as shown in Fig.~\ref{fig_iFactor}(c), the CDFs with different $\eta$ converge to the same, where the probability of unfulfilled rate requirement is nearly $0$, and the majority of the actual user rates $r_{k,u}$ fall into $[\gamma_{k,u},1.2\gamma_{k,u}]$. Similarly, in Fig.~\ref{fig_iFactor}(d), the energy efficiencies with different $\eta$ also converge in the $10$th period, despite their difference in the $1$st period. Therefore, the convergence of our proposed scheme under various $\eta$ is verified.

\begin{figure*}[tp]
 \begin{center}
 \includegraphics[width=1.75\columnwidth]{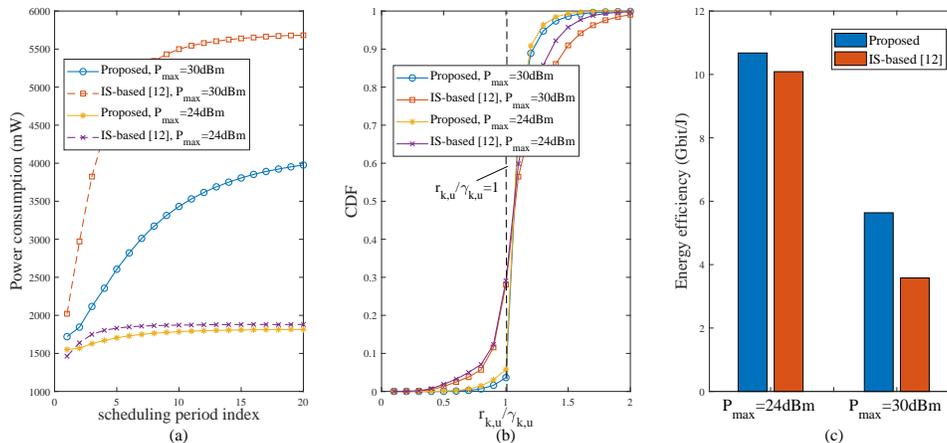}
 \end{center}
 \vspace{-5mm}
 \caption{Performance comparison in consecutive scheduling periods over $1000$ network realizations with $\eta=0.8$. (a) Total transmit power; (b) CDF of $\frac{r_{k,u}}{\gamma_{k,u}}$ in the $10$th scheduling period; (c) Energy efficiency in the $10$th scheduling period.}
 \vspace{-4mm}
 \label{fig_consecutive}
\end{figure*}

\begin{figure*}[tp]
	\begin{center}
		\includegraphics[width=1.75\columnwidth]{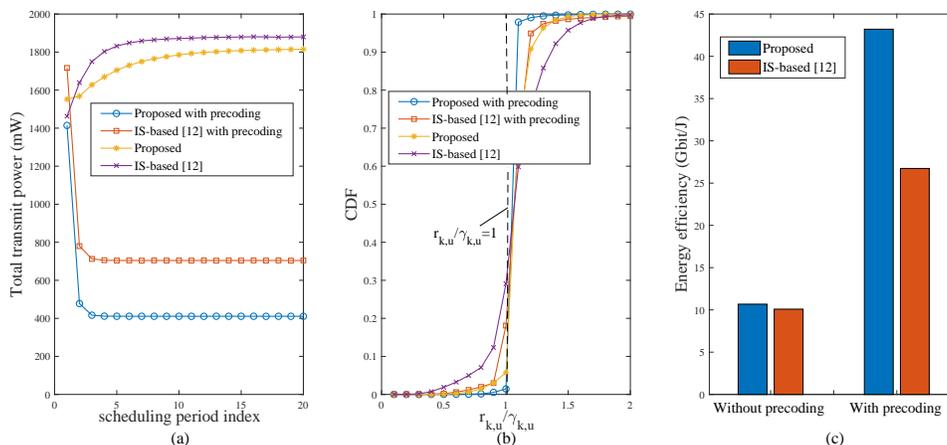}
	\end{center}
	\vspace{-5mm}
	\caption{Performance comparison with ZF precoding over $1000$ network realizations with $\eta=0.8$. (a) Total transmit power; (b) CDF of $\frac{r_{k,u}}{\gamma_{k,u}}$ in the $10$th scheduling period; (c) Energy efficiency in the $10$th scheduling period.}
	\vspace{-5mm}
	\label{fig_precoding}
\end{figure*}

 Then, we compare the performance of our proposed scheme with \cite{mmWave1} over $1000$ random network realizations in Fig.~\ref{fig_consecutive}, where $P_{max}=24,30$dBm, respectively. We consider $P_{max}=24$dBm firstly. Although the total transmit power of our scheme is slightly lower compared with \cite{mmWave1}, the proportion of unfulfilled rate requirements in the $10$th period of our scheme could be much lower. Besides, the actual rates of a large proportion of users in scheme \cite{mmWave1} fall into $[1.3\gamma_{k,u},2\gamma_{k,u}]$ which exceeds the required rate excessively, while the corresponding proportion of users in our scheme is very small. Hence, more energy is wasted by \cite{mmWave1} to serve the already fulfilled users, leaving the rate requirements of many users unfulfilled. Next, when $P_{max}$ increases to $30$dBm, the total transmit power of both two schemes boost drastically, but the proportion of unfulfilled rate requirements decreases slightly. This phenomenon results from serving the users with poor channel condition and high rate requirement, which would be extremely power-consuming. Since the rate requirements of these users might be fulfilled after $P_{max}$ increases, the total transmit power grows significantly and the energy efficiency decreases. In this case, our scheme has much less transmit power and unfulfilled requirements compared with \cite{mmWave1}, and obtain an obvious performance gain on energy efficiency.

\subsection{Effects of Tx Digital Precoding}\label{sec_sim_precoding}
 Finally, the performance with Tx digital precoding is evaluated. At each time slot, we consider that all BSs perform ZF digital precoding for the $N_{RF}$ local users served in SDMA manner, where the channel coefficient of the equivalent baseband channel from a user to a BS is generated by the channel gain after beamforming with a random phase in $[0,2\pi]$. Since all strong interferences
 are avoided in NIF scheduling, the intra-cell interferences from the local BS could become dominant in the total interference power because of the relatively close distance. On the other hand, ZF precoding is able to cancel the intra-cell interference, leaving only inter-cell interference affecting the user SINR. Therefore, the network performance could be enhanced significantly.

 In Fig.~\ref{fig_precoding}, the network performance with ZF precoding is compared with the performance without ZF precoding, where $P_{max}=24$dBm. As expected, since the dominant intra-cell interferences in the residual interferences are cancelled by ZF precoding, a remarkable performance gain could be derived for both the scheme in \cite{mmWave1} and our scheme, in terms of all metrics in Fig.~\ref{fig_precoding}(a)-(c). Besides, the convergence of network performance also becomes faster due to the lowered interference level. The performance gain of our scheme compared with \cite{mmWave1} becomes more significant when ZF precoding is adopted. This is because the requirements of more users with poor channel condition can be fulfilled owing to the precoding technique, leading to a higher probability of feasibility of our scheme to fully exploit the performance gain brought by our proposed NIF scheduling. Especially, it can be seen from Fig.~\ref{fig_precoding}(b) that the probability of unfulfilled user in our scheme almost decreases to $0$ after ZF precoding.

\section{Conclusions}\label{sec_conclusion}
 In this paper, the NIF space-time user scheduling in mmWave cellular network is studied. Firstly, we investigate the user scheduling problem to minimize the unfulfilled user requirements on resource elements under NIF condition. A clique-based lower bound of the unfulfilled requirements on resource elements is derived, and a near-optimal scheduling algorithm is proposed accordingly. Simulation results verify that our proposed algorithm almost ensures zero unfulfilled requirement if the requirement $d_{k,u}$ on resource elements is well-designed such that the clique-based lower bound is zero. Furthermore, we study the joint user scheduling and power allocation problem under NIF condition, where the total transmit power is minimized to fulfill the user requirements on data rate. Based on the proposed NIF scheduling, a near-optimal joint user scheduling and power allocation scheme is designed, which achieves superior performance with limited CSI when compared with its conventional counterpart.

\bibliographystyle{IEEEtran}

\begin{thebibliography}{10}

\bibitem{background1}
 R. W. Heath, \emph{et al.}, ``An overview of signal processing techniques for millimeter wave MIMO system,'' \emph{IEEE J. Sel. Topics Signal Process.}, vol.~10, no.~3, pp.~436--453, Apr. 2016.

\bibitem{background2}
 O. E. Ayach, \emph{et al.}, ``Spatially sparse precoding in millimeter wave MIMO systems,'' \emph{IEEE Trans. Wireless Commun.}, vol.~13, no.~3, pp.~1499--1513, Mar. 2014.

\bibitem{background3}
 M. Xiao, \emph{et al.}, ``Millimeter Wave Communications for Future Mobile Networks,'' \emph{IEEE J. Sel. Areas Common.}, vol.~35, no.~9, pp.~1909--1935, Sep. 2017.

\bibitem{background4}
 S. Kutty, and D. Sen, ``Beamforming for Millimeter Wave Communications: An Inclusive Survey,'' \emph{IEEE Commun. Surveys Tuts.}, vol.~18, no.~6, pp.~949--973, Dec. 2015.

\bibitem{background5}
 M. Giordani, M. Mezzavilla, and M. Zorzi, ``Initial access in 5G mmWave cellular networks,'' \emph{IEEE Commun. Mag.}, vol.~54, no.~11, pp.~40--47, Nov. 2016.

\bibitem{background6}
 K. Ma, P. Zhao, and Z. Wang, ``Deep learning assisted beam prediction using out-of-band information,'' in \emph{Proc. 2020 IEEE 91st Vehicular Technology Conference (VTC2020 Spring)}, Antwerp, Belgium, May 2020.
 
\bibitem{interference1}
 S. Singh, R. Mudumbai, and U. Madhow, ``Interference Analysis for Highly Directional 60-GHz  Mesh Networks: The Case for Rethinking Medium Access Control,'' \emph{IEEE/ACM Trans. Netw.},  vol.~19, no.~5, pp.~1513--1527, Oct. 2011.
 
\bibitem{interference2}
 M. D. Renzo, ``Stochastic Geometry Modeling and Analysis of Multi-Tier Millimeter Wave Cellular Networks,'' \emph{IEEE Trans. Wireless Commun.}, vol.~14, no.~9, pp.~5038--5057, Sep. 2015.
 
\bibitem{interference3}
 M. Rebato, \emph{et al.}, ``Understanding Noise and Interference Regimes in 5G Millimeter-Wave Cellular Networks,'' in \emph{Proc. 22th European Wireless Conference}, Oulu, Finland, May 2016.

\bibitem{low1}
 G. Y. Li, \emph{et al.}, ``Multi-Cell Coordinated Scheduling and MIMO in LTE,'' \emph{IEEE Commun. Surveys Tuts.}, vol.~16, no.~2, pp.~761--775, Mar. 2014.

\bibitem{low2}
 E. Castaneda, A. Silva, A. Gameiro, and M. Kountouris, ``An Overview on Resource Allocation Techniques for Multi-User MIMO Systems,'' \emph{IEEE Commun. Surveys Tuts.}, vol.~19, no.~1, pp.~239--284, Oct. 2016.

\bibitem{mmWave1}
 Y. Niu, \emph{et al.}, ``Energy-Efficient Scheduling for mmWave Backhauling of Small Cells in Heterogeneous Cellular Networks,'' \emph{IEEE Trans. Veh. Technol.}, vol. 66, no. 3, pp.~2674--2687, Mar. 2017.

\bibitem{mmWave2}
 B. Fan, \emph{et al.}, ``A Cross-Tier Scheduling Scheme for Multi-Tier Millimeter Wave Wireless Networks,'' \emph{IEEE Trans. Wireless Commun.}, vol.~17, no.~8, pp.~5029--5044, Aug. 2018.

\bibitem{mmWave3}
 X. Qin, \emph{et al.},``Joint User-AP Association and Resource Allocation in Multi-AP 60-GHz WLAN,'' \emph{IEEE Trans. Veh. Technol.}, vol.~68, no.~9, pp.~5696--5710, Jun. 2019.

\bibitem{mmWave4}
 Z. Sha, and Z. Wang, ``Least Pair-Wise Collision Beam Schedule for mmWave Inter-Cell Interference Suppression,'' \emph{IEEE Trans. Wireless Commun.}, vol.~18, no.~9, pp.~4436-4449, Sep. 2019.

\bibitem{mmWave5}
 Z. Sha, Z. Wang, S. Chen, and L. Hanzo, ``Graph Theory Based Beam Scheduling for Inter-Cell Interference Avoidance in MmWave Cellular Networks,'' \emph{IEEE Trans. Veh. Technol.}, vol.~69, no.~4, pp.~3929--3942, Apr. 2020.

\bibitem{mmWave6}
 Z. Zhou, W. Feng, Y. Chen, and N. Ge, ``Adaptive scheduling for millimeter wave multi-beam satellite communication systems,'' \emph{J. Commun. Inf. Netw.}, vol.~1, no.~3, pp.~42--50, Oct.~2016.

\bibitem{mmWave0}
 C. Sum, and H. Harada ``Scalable Heuristic STDMA Scheduling Scheme for Practical Multi-Gbps Millimeter-Wave WPAN and WLAN Systems,'' \emph{IEEE Trans. Wireless Commun.}, vol.~11, no.~7, pp.~2658--2669, Jul. 2012.

\bibitem{low3}
 J. Zhang, \emph{et al.}, ``Networked MIMO with Clustered Linear Precoding,'' \emph{IEEE Trans. Wireless Commun.}, vol.~8, no.~4, pp.~1910--1921, Apr. 2009.

\bibitem{low4}
 N. Seifi, \emph{et al.}, ``Coordinated 3D Beamforming for Interference Management in Cellular Networks,'' \emph{IEEE Trans. Wireless Commun.}, vol.~13, no.~10, pp.~5396--5410, Oct. 2014.

\bibitem{low5}
 S. Moon, C. Lee, S. R. Lee, and I. Lee, ``Joint User Scheduling and Adaptive Intercell Interference Cancelation for MISO Downlink Cellular Systems,'' \emph{IEEE Trans. Veh. Technol.}, vol.~62, no.~1, pp.~172--181, Jan. 2013.

\bibitem{low6}
 F. Wang, W. Chen, H. Tang, and Q. Wu, ``Joint Optimization of User Association, Subchannel Allocation, and Power Allocation in Multi-Cell Multi-Association OFDMA Heterogeneous Networks,'' \emph{IEEE Trans. Commun.}, vol.~65, no.~6, pp.~2672--2684, Jun. 2017.

\bibitem{low7}
 W. Yu, T. Kwon, and C. Shin, ``Multicell Coordination via Joint Scheduling, Beamforming, and Power Spectrum Adaptation,'' \emph{IEEE Trans. Wireless Commun.}, vol.~12, no.~7, pp.~3300--3313, Jul. 2013.

\bibitem{low8}
 F. Riera-Palou, and G. Femenias, ``Cluster-Based Cooperative MIMO-OFDMA Cellular Networks: Scheduling and Resource Allocation,'' \emph{IEEE Trans. Veh. Technol.}, vol.~67, no.~2, pp.~1202--1216, Feb. 2018.

\bibitem{mmWave01}
 Z. Xiao, L. Zhu, and X. Xia, ``UAV Communications with Millimeter-Wave Beamforming: Potentials, Scenarios, and Challenges,'' \emph{China Commun.}, vol.~17, no.~9, pp.~147-166, Sep. 2020.

\bibitem{mmWave02}
 L. Zhu, J. Zhang, Z. Xiao, X. Cao, D. O. Wu, and X. Xia, ``Millimeter-Wave NOMA with User Grouping, Power Allocation and Hybrid Beamforming,'' \emph{IEEE Trans. Wireless Commun.}, vol.~18, no.~11, pp.~5065-5079, Nov. 2019.

\bibitem{mmWave7}
 Y. Xu, H. S. Ghadikolaei, and C. Fischione, ``Adaptive Distributed Association in Time-Variant Millimeter Wave Networks,'' \emph{IEEE Trans. Wireless Commun.}, vol.~18, no.~1, pp.~459--472, Jan. 2019.

\bibitem{mmWave8}
 W. Feng, \emph{et al.}, ``When mmWave Communications Meet Network Densification: A Scalable Interference Coordination Perspective,'' \emph{IEEE J. Sel. Areas Common.}, vol.~35, no.~7, pp.~1459--1471, Jul. 2017.

\bibitem{mmWave9}
 B. Soleimani, and M. Sabbaghian, ``Cluster-Based Resource Allocation and User Association in mmWave Femtocell Networks,'' \emph{IEEE Trans. Commun.}, vol.~68, no.~3, pp.~1746-1759, Mar. 2020.

\bibitem{IS1}
 K. Yang, D. Calin, C.-B. Chae, and S. Yiu, ``Distributed Beam Scheduling in Multi-cell Networks via Auction over Competitive Markets'', in \emph{2011 IEEE International Conference on Communications (ICC)}, Kyoto, Japan, Jul. 2011.

\bibitem{IS2}
 K. Ahuja, Y. Xiao, and M. v. d. Schaar, ``Distributed Interference Management Policies for Heterogeneous Small Cell Networks,'' \emph{IEEE J. Sel. Areas Common.}, vol.~33, no.~6, pp.~1112--1126, Jun. 2015.

\bibitem{IS3}
 K. Ahuja, Y. Xiao, and M. v. d. Schaar, ``Efficient Interference Management Policies for Femtocell Networks,'' \emph{IEEE Trans. Wireless Commun.}, vol.~14, no.~9, pp.~4879--4893, Sep. 2015.

\bibitem{IS4}
 I. C. Paschalidis, F. Huang, and W. Lai, ``A Message-Passing Algorithm for Wireless Network Scheduling,'' \emph{IEEE/ACM Trans. Netw.}, vol.~23, no.~5, pp.~1528--1541, Oct. 2015.

\bibitem{IS5}
 C. Joo, X. Lin, J. Ryu, and N. B. Shroff, ``Distributed Greedy Approximation to Maximum Weighted Independent Set for Scheduling With Fading Channels,'' \emph{IEEE/ACM Trans. Netw.}, vol.~24, no.~3, pp.~1476--1488, Jun. 2016.

\bibitem{Proofs}
 Z. Sha, S. Chen, and Z. Wang, ``Supplementary Materials for Article: Near Interference-Free Space-Time User Scheduling for MmWave Cellular Network,'' \emph{Researchgate Preprint}, available: https://www.researchgate.net/publication/353117005.

\bibitem{3GPP}
 3GPP, TR38.900, ``Study on channel model for frequency spectrum above 6 GHz (Release 15),'' V15.0.0, Jun. 2018.

\bibitem{convex}
 S. Boyd, and L. Vandenberghe, \emph{Convex Optimization}, Cambridge, U.K., Cambridge Univ. Press, 2004.

\end{thebibliography}

\end{document}